\documentclass[aps,pre,preprint,groupedaddress]{revtex4}
\usepackage{graphicx}
\usepackage{amssymb,amsfonts,amsmath}
\usepackage{mathrsfs}
\usepackage{color}
\usepackage[3D]{movie15}
\usepackage{hyperref}
\usepackage{bm}
\usepackage{ulem}
\MakeRobust{\overrightarrow}

\usepackage{epstopdf}
\newif\ifpdf
\ifx\pdfoutput\undefined
   \pdffalse
\else
   \pdfoutput=1
   \pdftrue
\fi
\ifpdf
   \usepackage{graphicx}
   \usepackage{epstopdf}
\else
   \usepackage{graphicx}
\fi

\begin{document}

\title{Elastocapillary coalescence of plates and pillars}

\author{Z. WEI$^1$, T.M. SCHNEIDER$^{1,4}$, J. KIM$^5$,  H.-Y. KIM$^5$, J. AIZENBERG$^{1,3}$ \\and L. MAHADEVAN$^{1,2,3}$}
\email{lm@seas.harvard.edu}

\affiliation{$^1$School of Engineering and Applied Sciences, $^2$Department of Physics, $^3$ Kavli Institute for Nano-Bio Science and Technology, Wyss Institute for Bio-inspired Engineering, Harvard University, Cambridge, Massachusetts 02138, USA\\$^4$Max Planck Institute for Dynamics and Self-Organization, Am Fassberg 17, D-37077 Goettingen, Germany
\\$^5$Department of Mechanical and Aerospace Engineering, Seoul National University, Seoul 151-744, Korea}

\begin{abstract}
When a fluid-immersed array of lamellae or filaments that is attached to a substrate is dried, evaporation leads to the formation of menisci on the tips of the plates or pillars that bring them together.  Similarly, when hair dries it clumps together due to capillary forces induced by the liquid menisci between the flexible hairs. Building on prior experimental observations, we use a combination of theory and computation to understand the nature of this instability and its evolution in both the two-dimensional and three-dimensional setting of the problem. For the case of lamellae, we explicitly derive the interaction torques based on the relevant physical parameters. A Bloch-wave analysis for our periodic mechanical system captures the critical volume of the liquid and the 2-plate-collapse eigenmode at the onset of instability.  We study the evolution of clusters and their arrest using numerical simulations to explain the hierarchical cluster formation and characterize the sensitive dependence of the final structures on the initial perturbations. We then generalize our analysis to treat the problem of pillar collapse in 3D, where the fluid domain is completely connected and the interface is a  surface with the uniform mean curvature. Our theory and simulations capture the salient features of both previous experimental observations and our own in terms of the key parameters that can be used to control the kinetics of the process.
\end{abstract}

\maketitle

\section{Introduction}\label{sec:intro}
While assembly of complex macromolecular structures on microscopic lengthscales \cite{Desiraju1989} is driven by van der Waals interactions, dispersive forces and chemical interactions between constituents, on mesoscopic length scales of the order of microns to millimeters in the context of colloids and larger particles, other surface forces such as those due to capillarity play an important role \cite{Bowden1997,Rothemund2000,Vella2005}.  When particles interact with each other via capillary forces, the resulting patterns are a function of the size and shape of the constituents, and any constraints on their movement. Capillary coalescence is a natural consequence of this and occurs when free particles aggregate at an interface \cite{Grzybowski2000,Grzybowski2001}, and also when extended objects such as filaments and lamellae are brought together by interfacial forces which drive aggregation \cite{Pokroy2009,Bico2004, Kim2006,Tanaka1993}. Sometimes, these systems coarsen indefinitely leading to a single cluster, while at other times  elastic deformations eventually arrest the process leading to many finite sized clusters.  In a typical experimental setting characterizing the latter, fibers, filaments or lamellae are fully immersed in a liquid which is subsequently evaporated so that capillary forces at the liquid-gas interfaces bring the constituents together. Fibers and lamellae are long and soft objects and can easily bend. Thus a competition between actuating capillarity and resisting elasticity selects the structures formed,  which can be used to construct substrates with tunable wetting and adsorption properties \cite{Bernardino2010,Bernardino2012}. A variety of experimental systems that fall into this category include millimeter-scaled macroscopic brush hairs \cite{Bico2004}, micrometer-scaled mesoscopic polymeric surface mimicking gecko foot hairs \cite{Geim2003}, as well as nanometer-scaled carbon nanotube forests \cite{Lau2003, Chakrapani2004}.

Elastocapillary interaction have been well characterized in two-body systems \cite{Kim2006, Duprat2011, Taroni2012}. However, for the collective behavior of many elastic fibers bundling together, typically static energy minimization arguments have been employed to estimate the expected finally assembled bundle size \cite{Bico2004, Chandra2009, Zhao2006}, although recent work \cite{Gat2013} uses stability analysis of the unpatterned base state to predict the ordered hierarchical coalescence structure of an array of clamped parallel elastic sheets which are partially immersed in liquid. Complementing these discrete approaches, a phenomenological continuum mean-field approach has been used to model the arrested coarsening \cite{Boudaoud2007}, while a recent continuum theory based on microscopic physics for the coupled dynamics of drying and coalescence explains the kinetics and refinement seen in experimental studies \cite{Wei2014}. Unlike in ergodic systems where the state space is stochastically sampled due to thermal excitations, the structures in elastocapillary systems are often not selected by energetics alone. Instead, they depend critically on the dynamics of the drying process. This leads to a path dependence caused by the strong coupling of the geometry of the air-liquid interface to the local evaporation when multiple unconnected liquid domains are formed. Additional aspects, such as pinning and contact angle hysteresis, as well as the permanent adhesion of contacts formed in intermediate structures, make purely energetic arguments unable to have any predicative power. Thus, to predict and control the assembled structures in capillarity-driven self-assembly experiments, we need to (1) follow the dynamics over time as irreversible effects associated with evaporation, contact line motion and adhesion, and (2) account for non-linear effects due to large deviations from the background state and go beyond a linearized analysis.

\begin{figure}
  \centering
  \includegraphics[width=1\textwidth]{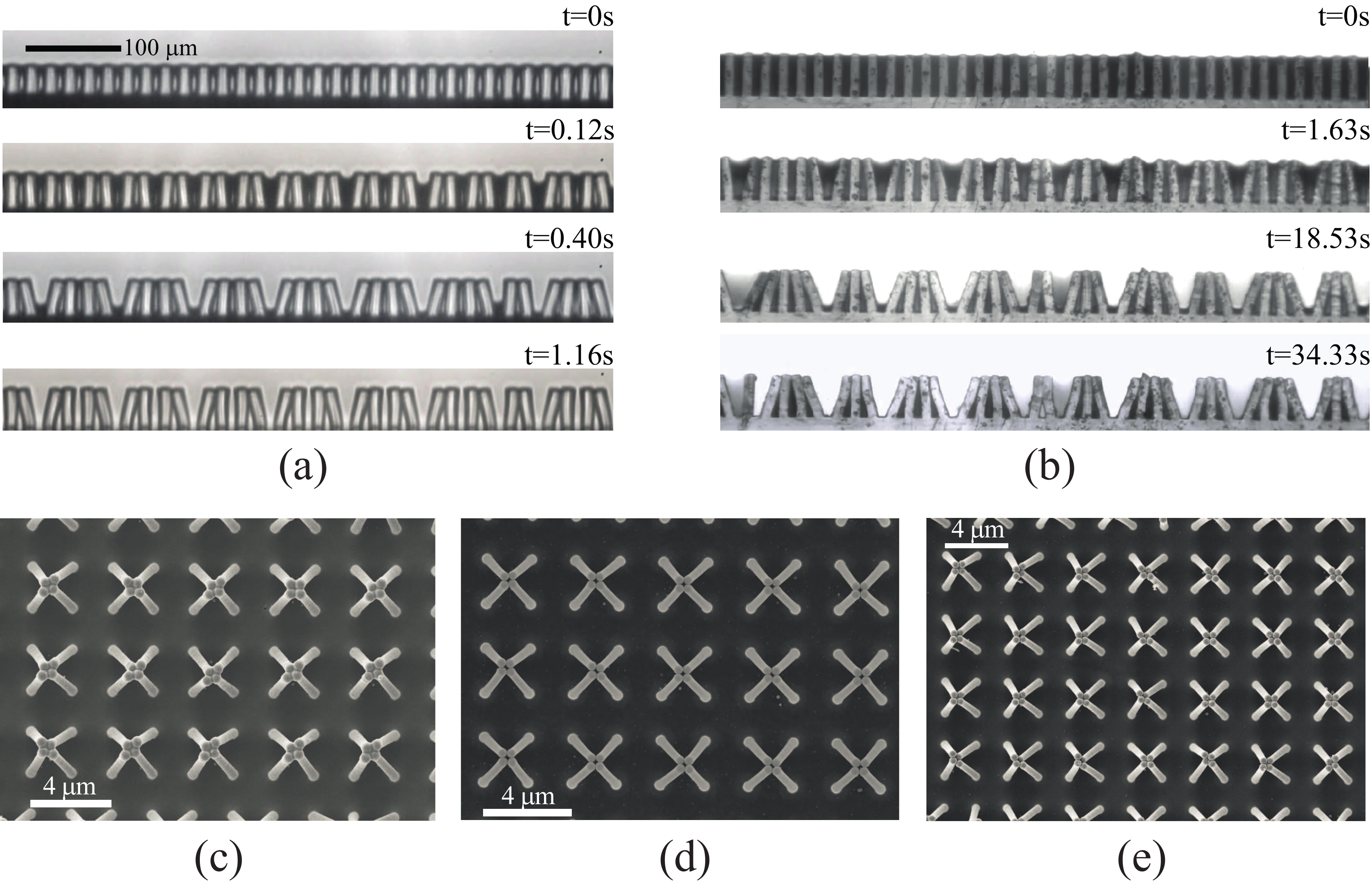}
  \caption{Dynamics of coalescence of arrays of (a-b) lamellae and (c-e) filaments driven by drying. (a) The lamellae are of { thickness $h=10\mu m$, depth $W=40\mu m$  and height $L=40 \mu m$, and spaced a distance $D=10\mu m$ apart}. As the liquid evaporates, a front of dimerization coalescence initiates and propagates from right to left, after which a front of quadrimerization moves through the system in the same direction. After the liquid dries out, the quad-bundles separate and the dimers persist. (b) {The lamellae are of depth $\sim 1mm$, and other geometric parameters are the same as those in (a).} Due to the imperfections in the system, e.g. the lamella geometry, roughness of the surface and et al., coalescence initiates simultaneously here. The 2-lamella mode still appears to be dominant at the onset of instability, and irregular bundles with a size distribution from 2-5 arise thereafter and persist even after the liquid dries out. (c) The final structures of the two-dimensional array of pillars of diameter after the liquid dries out form four-fold rhombic clusters. (d) Here the tips of the fourfold clusters form squares. (e) A larger domain shows that the four fold clusters are more asymmetric, and the tips form both rhombi and squares. The geometric parameters of the pillars and the material properties are the same in all three images; the pillars have a diameter $300 nm$, height $4.5\mu m$ and spacing $2\mu m$. The only difference is the size of imperfections in  the system.}
  \label{experiments}
\end{figure}

Here we use a combination of theory and computation to understand the instability and the hierarchical evolution of cluster formation for both two- and three-dimensional elastocapillary systems driven by drying. Those systems represent generic situations, allow for a thorough theoretical treatment and can be validated by well controlled experiments. For the two dimensional case, we describe the drying induced collapse of an one-dimensional array of evenly spaced lamellae immersed in the evaporating liquid (Figure \ref{experiments}(a)(b)). For this system all relevant physical forces are considered, allowing us to derive the interaction potentials and forces, analyze the linear stability of the system, and compute the nonlinear dynamics associated with pattern coarsening. We demonstrate that different dynamical paths through the system's state space indeed lead to different final structures which are observed in experiments.

In the three-dimensional case, our theory focuses on explaining experiments associated with the bundling of a regular square grid of fluid-immersed elastic posts anchored to a substrate \cite{Pokroy2009} (Figure \ref{experiments}(c)-(e)). We determine the constant mean curvature surface for the air-liquid interface subjected to the global liquid volume constraint on the multi-connected domain. This allows us to compute the interaction forces and thence the primary unstable mode. Finally we use a numerical method to compute the aggregation dynamics.

In $\S$\ref{sec:2d} we describe the experimental observations for the two-dimensional case, derive a discrete two-plate model for the deformation of the plates driven by capillary forces, and carry out a linear stability analysis of the base state. We then study the coalescence dynamics of a collection of plates in the nonlinear regime. In $\S$\ref{sec:3d} we describe the experimental observations for the three-dimensional case, and introduce a dynamical model that allows us to simulate the morphology of clustered pillars, and compare these results with experiments. In $\S$\ref{sec:conclusion}, we conclude with a description of open problems in this rich area.

\section{Collective dynamics of elastic lamellae}\label{sec:2d}
\subsection{Experimental observations}\label{subsec:2dexp}
For the two-dimensional case (Figure \ref{model}), we consider a one-dimensional array of elastic micro-lamellae with height $L$, thickness $h$ and uniform spacing $D$, Young's modulus $E$ and Poisson's ratio  $\nu$ respectively.  Each lamella is assumed to be free at one end and anchored at the other on a substrate \cite{Tanaka1993}. The lamella array is wetted by a liquid of surface tension $\sigma$, density $\rho$ and viscosity $\mu$, which is confined between neighboring lamellae, defining a cell. The contact line slips from the tips as the liquid evaporates. When the system is completely immersed in the liquid, the stable configuration is a uniform array of non-interacting vertical lamellae. However, when the liquid evaporates, it is not necessarily locally stable any more: capillary forces associated with the liquid-air menisci between the free ends of the soft lamellae may cause them to deflect laterally and adhere together. In an experimental system with small imperfections, we observe a regular cascade of successive sticking events that leads to a hierarchical bundling pattern: every two neighboring lamellae incline towards each other to form a dimer first, which then collapses into quadrimers (Figure. \ref{experiments}(a)). The process repeats until the bending deformation induced elastic resistance eventually becomes large enough to prevent further coarsening. In the system with large imperfections, irregular bundles can and do arise but the 2-lamella-collapse mode still appears to be dominant right after the instability(Figure. \ref{experiments}(b)). After the liquid dries out, bundles separate if the adhesion in contact is not strong to counterbalance the elastic forces; else they persist. In our experiments, the liquid used is isopropyl alcohol (IPA), and the lamellae and the substrate are made of polydimethylsiloxane (PDMS). Throughout this entire section, we use the following experimental parameters $\sigma=0.022 N/m$, $\rho = 786Kg/m^3$, $\mu = 0.0196Pa\cdot s$, $E=1.5Mpa$, $\nu=0.5$, $h=10\mu m$, $L=40\mu m$, $D=10\mu m$, and gravity $g=9.8m/s^2$.

\subsection{Mechanics: coupling plate bending to fluid interface shape}\label{subsec:2dmodel}
\begin{figure}
  \centering
  \includegraphics[width=0.8\textwidth]{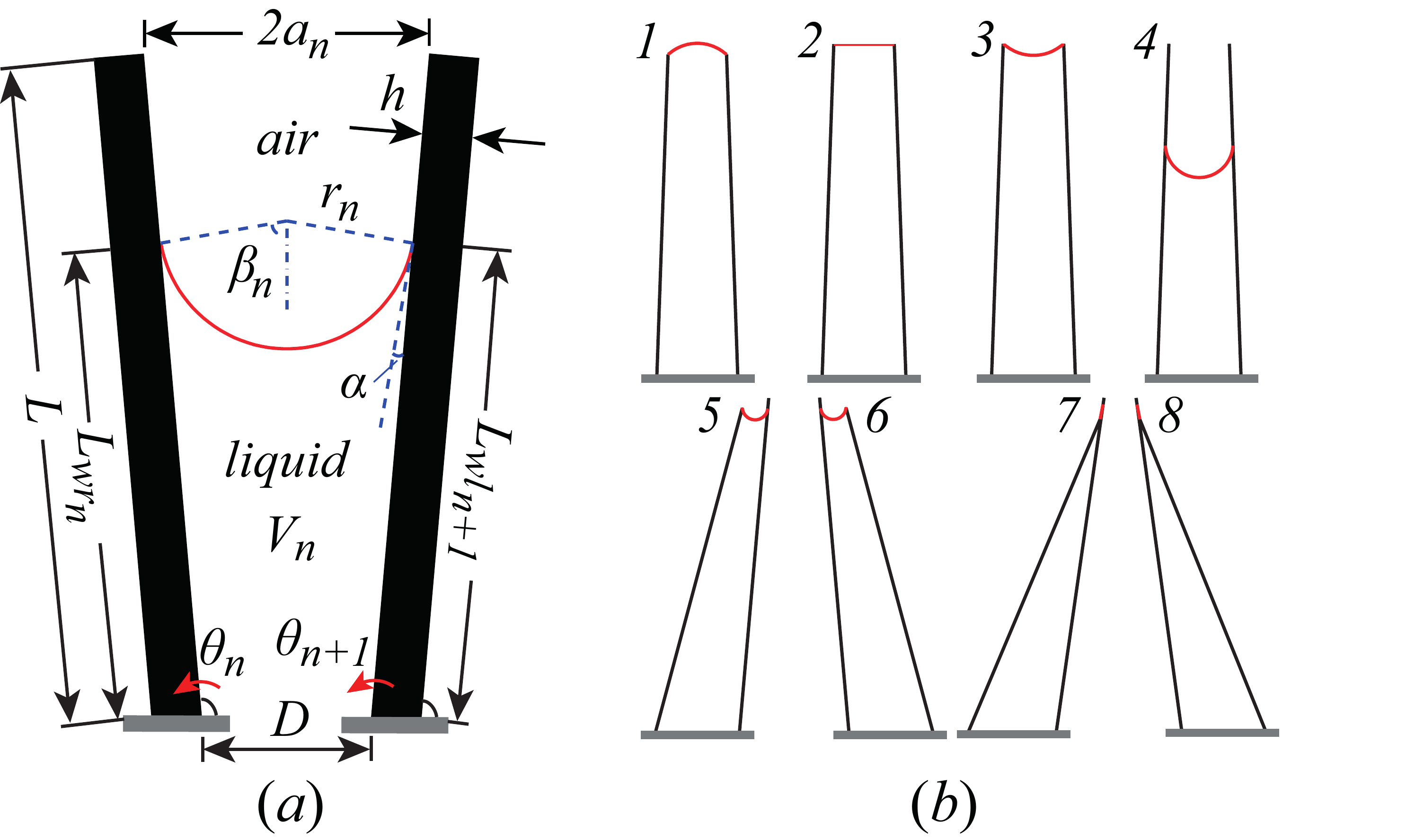}
  \caption{A unit cell confined by two adjacent lamellae/plates. (a)  Each lamella is modeled as a rigid plate elastically hinged at the base. The plate is of height $L$ and thickness $h$. $\theta_n$ is the angle of $n^{th}$ plate with respect to the horizontal. The plate spacing is $D$ and the distance between the two free tips is $2a_n$. $V_n$ is the volume per unit depth of the liquid confined in the cell. The red curve represents the air-liquid interface, an arc of a circle of radius $r_n$, with  $\beta_n$ being the half angle subtended by the meniscus arc. $\alpha$ is the critical contact angle at which the meniscus slides down from the tip. ${L_{wl}}_{n+1}$ is the wetting length on the left side of the $(n+1)^{th}$ plate and ${L_{wr}}_n$ is the wetting length on the right side of the $n^{th}$ plate. (b) 1-8 are the 8 possible cases of menisci, showing that the menisci can be pinned on both tips, or slip down from one or both tips. 7 and 8 show that when the two free tips are so close that no circular arcs exist to connect them, in which case the arc is replaced by a line. These situations are solely used to prevent simulation failure in rare cases.}
  \label{model}
\end{figure}

In our experiments, the lamellae are short and stiff, and remain almost straight as they are deflected by capillary forces, bending primarily in the neighborhood of the base. Therefore, we can approximate each lamella as a rigid plate and integrate all the bending response into an elastic hinge at the base \cite{Bernardino2012}. This simplifies our analysis relative to the case that must account for the inhomogeneous bending and buckling (APPENDIX \ref{AppendixA}) of the individual lamella \cite{Gat2013, Neukirch2007}.  The hinge elastic constant can be approximately derived from the bending response of a short cantilever by a transverse force $F$ at its free end, which is given by
\begin{equation}
k  = \frac{d(F\delta)}{d\theta} = \left(\frac{3L}{Eh^3}+\frac{6}{7GhL}\right)^{-1},
\label{Eq:k}
\end{equation}
where $\delta = {4FL^3(1-\nu^2)}/{Eh^3} + {\alpha_sFL}/{Gh}$  is the deflection at the free end, $\theta$ is the tilting angle between the straight plate and the horizontal direction (Figure \ref{model}(a)), $G$ is the shear modulus, and $\alpha_s$ is the shear coefficient. The second term of the right hand side of Eq.(\ref{Eq:k}) is due to shear deformations in the so-called Timoshenko beam theory \cite{Timoshenko1972}, because the slender beam condition $L\gg h$ is violated in our experimental setup. We have taken $\nu=0.5$ and  $\alpha_s = 10(1+\nu)/(12+11\nu)$ as an approximation for the rectangular cross section.

To explicitly derive the torques due to capillarity, we consider a unit cell consisting of two plates, with liquid confined in between, and air outside (Figure \ref{model}(a)). The pressure field inside the liquid is nonuniform due to the effects of gravity and flow. However, in our system, gravity can be neglected, because the Bond number $Bo = \Delta \rho g D^2/\sigma\sim 10^{-5}\ll1$. Comparing the viscous moment due to flow $M_{\mu}\sim\mu L^5\dot{\theta}/D^3$ {(APPENDIX \ref{AppendixC})} with that caused by the surface tension $M_{\sigma}\sim\sigma L^2/D$, we find that $M_{\mu}/M_{\sigma}\sim10^{-3}\ll1$ for $\dot{\theta}\sim O(1)$, which indicates that pressure effects due to fluid flow can be neglected. Therefore, the air-liquid interface is always a segment of a circle because of the uniform pressure in each cell. For each plate, the moment results from both the line tension at the contact line, and the pressure gradient across the plate caused by the curvature difference of two neighboring menisci. To calculate the moment we need to find the wetting length, the contact angle, and the pressure difference across the air-liquid interface due to its local curvature. We assume the contact angle  can take any value equal to or larger than a critical value $\alpha$, consistent with the fact that the meniscus can be pinned at the tip of the lamella while being concave-up (Figure \ref{model}(b1)), flat (Figure \ref{model}(b2)), or concave-down (Figure \ref{model}(b3)). We assume that once the contact angle reaches the critical value $\alpha$, it remains constant as the meniscus starts to slide down from the tip (Figure \ref{model}(b4)-(b6)). Consequently, the meniscus profile and the resulting moment can be calculated for any given tilting angles $\theta_n$, $\theta_{n+1}$ and the liquid volume per unit depth $V_n$. Scaling lengths by $L$, volumes $V_n$ by $L^2$ and moments by $\sigma L$ leads to the following results for the different cases that we summarize here for our subsequent linear stability analysis (for details, see {APPENDIX \ref{AppendixB}}).

\begin{enumerate}
\item The meniscus is pinned on both tips (Figure \ref{model}(b1)). The half angle $\beta_n$ subtended by the meniscus arc is determined by solving
    \begin{equation}
    V_n = \frac{1}{2}d(\sin\theta_n + \sin\theta_{n+1})+\frac{1}{2}\sin(\theta_n-\theta_{n+1})-a_n^2(\beta_n\csc^2\beta_n - \cot\beta_n),
    \label{Eq:pin_V}
    \end{equation}
    for given $V_n$, $\theta_n$ and $\theta_{n+1}$, where $d=D/L$ and tip separation is
    \begin{equation}
    2a_n = \sqrt{2-2\cos(\theta_{n+1}-\theta_n)+2d(\cos\theta_{n+1}-\cos\theta_n)+d^2}.
    \label{Eq:pin_a}
    \end{equation}
    $\beta_n$ must satisfy $\beta_n\leq\hat{\beta}_n$, where $\hat{\beta}_n$ is the critical angle at which the meniscus starts to slide down from at least one lamella. $\beta_n<0$ when the meniscus concaves down, $\beta_n=0$ when the meniscus is flat, and $\beta_n>0$ when the meniscus concaves up. The moments on the $n^{th}$ and $n+1^{th}$ plates are given respectively by
    \begin{align}
    \label{Eq:pin_M1}
    &M_n = -\frac{1}{2a_n}\left[\sin(\beta_n+\theta_n-\theta_{n+1})+d\sin(\beta_n+\theta_n)\right],\\
    \label{Eq:pin_M2}
    &M_{n+1} = \frac{1}{2a_n}\left[\sin(\beta_n+\theta_n-\theta_{n+1})-d\sin(\beta_n-\theta_{n+1})\right],
    \end{align}

\item The meniscus has slipped down from both tips (Figure \ref{model}(b3)). The contact angle is fixed at $\alpha$. When $\theta_n=\theta_{n+1}=\theta$, the meniscus radius is independent of $V_n$, and $r_n=d\sin\theta/(2\cos\alpha)$. $l_n$ is determined by solving
\begin{equation}
V_n = \frac{2l_n-d\cos\theta}{2}d\sin\theta - \left(\frac{d\sin\theta}{2}\right)^2\tan\alpha - \frac{\pi-2\alpha}{2}\left(\frac{d\sin\theta}{2\cos\alpha}\right)^2.
\label{Eq:bothdown_V2}
\end{equation}
The wetting length on the right side of the $n^{th}$ plate and that on the left side of the $(n+1)^{th}$  plate are given respectively by
\begin{align}
\label{Eq:bothdown_Lwr2}
&{L_{wr}}_n = l_n-\frac{d}{2}\sin\theta\tan\alpha,\\
\label{Eq:bothdown_Lwl2}
&{L_{wl}}_{n+1} = l_n-\frac{d}{2}\sin\theta\tan\alpha-d\cos\theta.
\end{align}
The moments on the $n^{th}$ and $(n+1)^{th}$ plate are given respectively by
\begin{align}
\label{Eq:bothdown_M1}
&M_n = -\frac{{L_{wr}}_n^2}{2r_n}-{L_{wr}}_n\sin\alpha,\\
\label{Eq:bothdown_M2}
&M_{n+1} = \frac{{L_{wl}}_{n+1}^2}{2r_n}+{L_{wl}}_{n+1}\sin\alpha.
\end{align}
\end{enumerate}

To get the total moment on a plate, we must add the contributions from  adjacent cells. For example, we can obtain the full expression of $M_n$ simply by replacing $n$ by $n-1$ in Eq. (\ref{Eq:pin_M2}) and add up to Eq. (\ref{Eq:pin_M1}) for the case when the meniscus is pinned at both tips, which readily yields $M_n = M_n(\theta_{n-1},\theta_n,\theta_{n+1},V_n,V_{n-1})$.

The dynamics of the $n^{th}$ plate neglecting inertia {(APPENDIX \ref{AppendixC})} follows the overdamped first order equation of motion
\begin{equation}
C\frac{\partial{\theta}_n}{\partial t}+k\left(\theta_n-\frac{\pi}{2}\right) + \sigma L M_n(\theta_{n-1},\theta_n,\theta_{n+1},V_n,V_{n-1})=0,
\label{Eq:1storder}
\end{equation}
where $C$ is the damping coefficient, $k$ is defined in Eq. (\ref{Eq:k}) and the dimensionless moment $M_n$ is due to capillarity. (We use Eqs. (B3) and (B4),  Eqs. (B11) and (B12), Eqs. (B17) and (B18), Eqs. (B23)-(B26), and Eqs. (B29) and (B30) for different situations to obtain the full expressions of $M_n$ as explained above. Please see {APPENDIX \ref{AppendixB}} for more details.) To estimate $C$, we need to account for both the internal viscosity of the solid and the external viscosity of the fluid, and find that the former dominates, which gives $C\approx \tau_m k$ {(APPENDIX \ref{AppendixC})}. Together with Eq. (\ref{Eq:1storder}) and the dynamics of drying  that will be discussed later, we can now understand the dynamics of the lamella array completely.

\subsection{Onset of bundling: linear stability of the uniform base state}\label{subsec:stability}
For the base state with all lamella being vertical and a uniform meniscus associated with constant liquid volume in each cell is decreased, there is a potential for instability as the curvature of the menisci increase.

\begin{figure}
  \centering
  \includegraphics[width=1.0\textwidth]{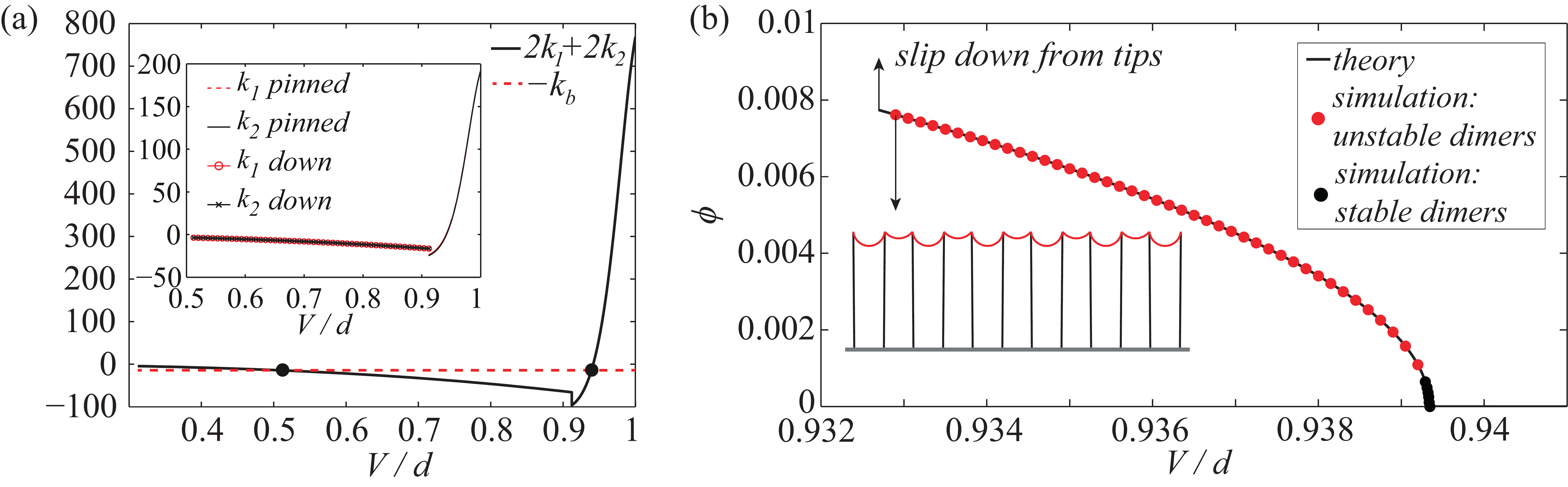}
  \caption{Stability analysis of a vertical array of plates. (a) The curve of $2k_1+2k_2$ following {Eqs. (\ref{Eq:pin_k1k2}) and (\ref{Eq:down_k1k2})} as a function of $V$ given by { Eqs. (\ref{Eq:pin_V_sym}) and (\ref{Eq:down_V})} intersects with the line $-k_b$. The interval between the two intersections indicates the region where the system is unstable as according to Eq. (\ref{Eq:eigenvectors}), $\lambda(\pi)=2k_1+2k_2+k_b<0$. For typical experiments, any $V/d\in(0.507,0.939)$ leads to coalesce given infinitesimal perturbations to the system. Inset figure shows the effective dimensionless spring constants $k_i(V)$ ($i=1, 2$); $k_i$ switches signs from positive to negative as $V$ decreases, while the discontinuity corresponds to the meniscus sliding down from the tip. (b) shows the positive bifurcation branch using both simulations of Eq. (\ref{Eq:1storder}) and the asymptotic result in Eq. (\ref{Eq:critical_dynamics}). When $V\ge V^*=0.9393d$, $\phi=0$, and when $V<V^*$, $\phi\sim \sqrt{V^*-V}$, which is a clear indication of supercritical bifurcation. The inset illustrates the dimer mode. We simulate 2 plates by solving Eq. (\ref{Eq:1storder}) with periodic boundary conditions, and display them for visualization. The black dots correspond to stable dimers, while the red dots correspond to unstable dimers that will further coalescence.}
  \label{super}
\end{figure}

To understand this, we consider a periodic domain of $2N$ plates and the same volume of liquid $V$ confined in each cell, when all plates being vertical ($\theta_n=\pi/2, n=1,2,...,2N$) is an equilibrium state. To determine the stability of this state, we study the perturbations in the moment as a function of variations in the angles $\theta_n$ linearized around the current state,
\begin{equation}
d\overrightarrow{M}=(\mathbb{K}_1+\mathbb{K}_2)d\overrightarrow{\theta},
\label{Eq:dM}
\end{equation}
where $\mathbb{K}_1$ and  $\mathbb{K}_2$ are the $2N\times2N$ stiffness matrices due to the elasticity of the plate and the geometrical change of menisci respectively. $\mathbb{K}_1$ is a  diagonal matrix with all elements being the dimensionless hinge constant $k_b =({3\sigma L^2}/{Eh^3}+{6\sigma}/{7Gh})^{-1}$, and $\mathbb{K}_2$ is a tridiagonal matrix with two additional elements of $-k_1$ on the upper-right and lower-left corners reflecting the periodic boundary condition. At the base state with $\theta_n=\pi/2$, $\mathbb{K}_2$ is expressed as
\begin{equation}
\mathbb{K}_2=\left[\begin{array}{ccccccc}
\ddots&\vdots&\vdots&\vdots&\vdots&\vdots &\\
\ldots&-k_1&2k_2&-k_1&0&0&\ldots \\
\ldots&0&-k_1&2k_2&-k_1&0&\ldots\\
\ldots&0&0&-k_1&2k_2&-k_1&\ldots \\
 &\vdots&\vdots &\vdots&\vdots &\vdots&\ddots
\end{array}\right],
\end{equation}
where $k_1=\partial M_n/\partial\theta_{n+1}=\partial M_{n+1}/\partial \theta_n$ and $k_2 = -\partial M_n/\partial\theta_n = -\partial M_{n+1}/\partial\theta_{n+1}$ are the stiffness of the effective spring connecting two neighboring plates due to capillarity, where $M_n$ is defined in Eq. (\ref{Eq:pin_M1}) or Eq. (\ref{Eq:bothdown_M1}), and $M_{n+1}$ is defined in Eq. (\ref{Eq:pin_M2}) or Eq. (\ref{Eq:bothdown_M2}). For the case when the meniscus is pinned at both tips,
\begin{equation}
\begin{split}
k_1 &=k_2+\sin\beta,\\
k_2 &= \frac{\cos\beta-d\sin\beta}{d}\left[1+\frac{1-d(\beta\csc^2\beta-\cot\beta)}{d^2\csc^2\beta(1-\beta\cot\beta)}\right]
-\frac{\sin\beta+d\cos\beta}{d^2},
\end{split}
\label{Eq:pin_k1k2}
\end{equation}
where $\beta$ is determined by the volume
\begin{equation}
V = d - (d/2)^2(\beta \csc^2 \beta -\cot \beta),
\label{Eq:pin_V_sym}
\end{equation}
which follows from Eq. (\ref{Eq:pin_V}) and Eq. (\ref{Eq:pin_a}) by substituting $\theta_n=\theta_{n+1}=\pi/2$. Here we have omitted the subscript $n$ for the translationally invariant base state. Note that $k_1$ and $k_2$ in this case are not necessarily equal, because while the wetting length is constant, the contact angles change by different amount on the two neighboring plates as they are deflected except when the meniscus is flat. For the case when the meniscus is no longer at both tips,
\begin{equation}
k_1 = k_2= -\left[\frac{L_w}{r}\sin\alpha + \frac{L_w^2}{2r^2}+\sin^2\alpha\right]\frac{2L_w + d\tan\alpha}{4\cos\alpha},
\label{Eq:down_k1k2}
\end{equation}
with $r = d/(2\cos\alpha)$ is the radius of the meniscus, and the wetting length $L_w$ is determined by the volume
\begin{equation}
V = L_w d+d^2\tan\left(\frac{\alpha}{4}\right) - d^2\frac{\pi-2\alpha}{8\cos^2\alpha},
\label{Eq:down_V}
\end{equation}
which follows from Eqs. (\ref{Eq:bothdown_V2})-(\ref{Eq:bothdown_Lwl2}) by substituting $\theta=\pi/2$. $k_1$ and $k_2$ in this case are always identical, because the contact angle keeps constant, and the change of wetting length is the same on both plates when they are deflected. From Eqs. (\ref{Eq:pin_k1k2})-(\ref{Eq:down_V}), we see that for any given geometric parameter $d$ and critical contact angle $\alpha$, $k_i=k_i(V)$ ($i=1, 2$). The inset of Figure \ref{super}(a) shows that $k_i$ switches sign as $V$ decreases, corresponding to the case when the system becomes unstable. We note that the discontinuity in $k_i$ corresponds to the transition from a static meniscus to a dynamic one that slides down from the tip.

To study the instability of the system, we investigate the eigenmodes of the stiffness matrix $\mathbb{K}=\mathbb{K}_1+\mathbb{K}_2$ which resembles a discrete Laplacian, with eigenvectors
\begin{equation}
\left[1, e^{if},e^{2if},\ldots,e^{i(2N-3)f},e^{i(2N-2)f},e^{-if}\right],
\label{Eq:eigenvectors}
\end{equation}
and the corresponding eigenvalues
\begin{equation}
\lambda(f) = -2k_1\cos(f)+2k_2+k_b, \ \text{where} \ f={n\pi}/{N}\ \text{and} \ n=1,2,\ldots,2N.
\label{Eq:eigenvalues}
\end{equation}
$\mathbb{K}$ must be positive definite to ensure stability, which is equivalent to requiring the smallest eigenvalue positive. When $k_1\ge0$, the meniscus is pinned at both tips, and the smallest eigenvalue is $\lambda(2\pi) = -2k_1+2k_2+k_b = k_b-2\sin\beta$, which follows from Eqs. (\ref{Eq:pin_k1k2}) and (\ref{Eq:eigenvalues}). In our experiments, $k_b\approx 13.5>2\sin\beta$, so that the array of vertical plates is stable. When $k_1<0$, the smallest eigenvalue is $\lambda(\pi) = 2k_1+2k_2+k_b$,  so that  stability is controlled  by the competition between elasticity and capillarity. As $k_i=k_i(V)$ ($i=1, 2$), the condition $\lambda(\pi)<0$ sets the range of $V$ in which the system is unstable (Figure \ref{super}(a)). The primary eigenmode corresponds to $f=\pi$, and the eigenvector thus is $[1,-1,1,-1,\ldots,1,-1]$ from Eq. (\ref{Eq:eigenvectors}), corresponding to dimerization of the lamellae. This calculation shows that to avoid the lamella collapse, we need to keep $k_1>0$ or reduce $|k_1|$ when $k_1<0$ and increase $k_b$. Practical approaches to implement this include the use of liquid with low surface tension and contact angle close to $90^o$, at which $k_1$ is at its maximum positive value, the use of stiff solids, and/or a proper choice of geometric parameters, e.g. a large aspect ratio $h/L$.

To explain the nature of the instability transition, we make use of the fact that the fastest growing mode is the dimer mode and assume $\theta_{n-1}-\pi/2 = \pi/2 - \theta_n = \theta_{n+1} - \pi/2 = \phi$. In the vicinity of the critical volume $V^*$ at which the instability happens, the dynamics of a lamella is governed by the equation
\begin{equation}
c\frac{d \phi}{d t} = -(2k_1+2k_2+k_b)\phi - g(V^*) \phi^3 +O(\phi^5),
\label{Eq:critical_dynamics}
\end{equation}
where $c$ is the dimensionless damping coefficient, and $g(V^*)>0$ is an algebraically lengthy coefficient of the cubic term. Eq. (\ref{Eq:critical_dynamics}) is derived by expanding Eq. (\ref{Eq:1storder}) in powers of $\phi$; even orders of $\phi$ do not appear in Eq. (\ref{Eq:critical_dynamics}) due to the reflection symmetry $\phi \to -\phi$ inherent in the system. When the volume in a cell reaches the critical value $V=V^*$ (equivalently $\beta=\beta^*$), $2k_1+2k_2+k_b=0$, while when $0<V^*-V\ll1$, a Taylor series expansion of Eq. (\ref{Eq:pin_k1k2}) in the neighborhood of $\beta^*$ gives us $2k_1+2k_2+k_b \sim \beta^*-\beta$, while a similar expansion of Eq. (\ref{Eq:pin_V_sym}) in the neighborhood of $\beta^*$ yields $V^*-V\sim \beta-\beta^*$, and hence $2k_1+2k_2+k_b \sim V-V^*$. From Eq. (\ref{Eq:critical_dynamics}), we see that the stable equilibrium state has two branches of solutions $\phi_1=-\phi_2\sim \sqrt{V^*-V}$ for $V^*>V$ and only one solution $\phi=0$ otherwise, suggesting that the bifurcation is supercritical. Figure \ref{super}(b) shows both the positive branch of the asymptotic solution $\phi_1$, and the results of simulation for $2$ lamellae obtained by solving Eq. (\ref{Eq:1storder}) with periodic boundary conditions. The results agree well, and confirm that the instability is supercritical and leads to lamellar dimerization, as shown in the inset.

\begin{figure}
  \centering
  \includegraphics[width=1\textwidth]{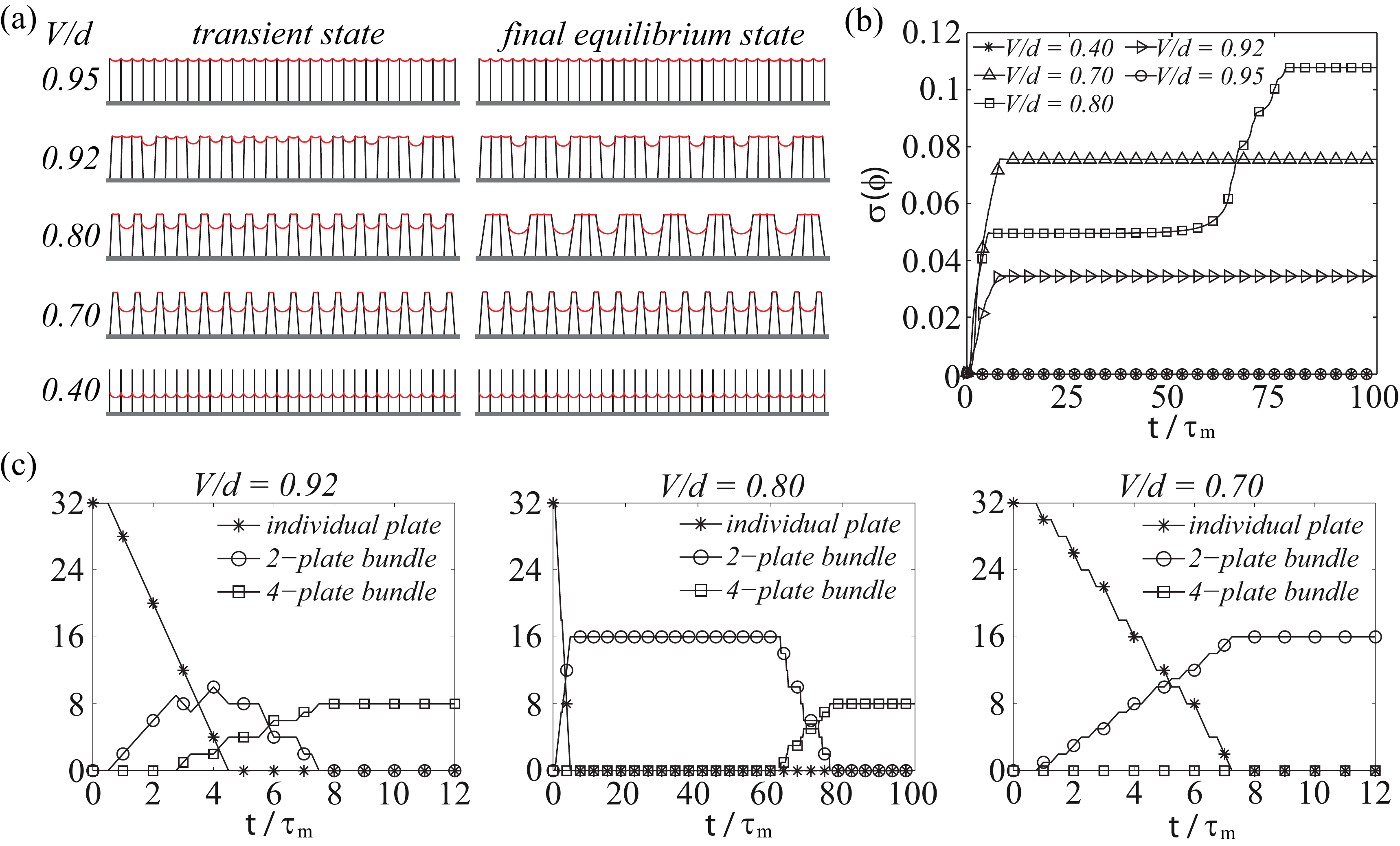}
  \caption{Elastocapillary coalescence of plates. (a) A periodic domain of 32 plates is simulated by solving Eq. (\ref{Eq:1storder}) for different prescribed volumes $V$. The tilting angle $\theta$ of the first plate is perturbed by $0.1\%$ from $90^o$ as the initial condition. The left column shows a sequence of  transient states and the right column shows the steady states. (b) The standard deviation of deflection angle $\phi$ as a function of time. $t$ is time and $\tau_m$ is the time scale for mechanical relaxation. (c) Complementing (b), these plots show the evolution of hierarchical formation of bundles for different prescribed volumes. The time resolution of the plots is $0.25\tau_m$ and plates are considered as a bundle if the tilting angle gradient is positive and larger than the perturbation amplitude.}
  \label{differentVs}
\end{figure}

\subsection{Nonlinear dynamics: drying, coarsening and refining}\label{subsec:2ddynamics}

\subsubsection{Controlled liquid volume}\label{subsubsec:2dvcontrol}
For an initially translationally invariant system, the two-plate-collapse mode is the fastest growing mode, yet the cluster of dimers is not necessarily the final stable state (red dots in Figure \ref{super}(b)). We notice that for a range of sufficient liquid volumes in each cell, both the deflection of the plates and the geometric change of the liquid menisci are large so that the linear approximation in Section \ref{sec:2d}(\ref{subsec:stability}) breaks down. Therefore, we numerically integrate Eq. (\ref{Eq:1storder}) directly to follow the hierarchical dynamics by which an array of vertical individual plates first forms dimers, then quadrimers, until eventually forming large bundles that are limited by elastic effects. Figure \ref{differentVs}(a) shows snapshots of dynamical coarsening for different control parameters $V$, in a periodic domain of 32 plates, triggered by tilting one plate by $0.1\%$ from $90^o$ as the initial condition. As expected, for volumes outside the range $V/d\in(0.5073,0.9393)$, e.g. for the cases of $V/d=0.95$ and $0.4$, the array is stable to perturbations and remains uniformly vertical. In the unstable parameter range, the primary mode corresponds to two plates collapsing into dimers (Figure \ref{super}(a)(c)). These dimers may further collapse into quadrimers or stay as the final stable state with different amplitudes of deformation angles as shown in Figure \ref{differentVs}(a)(b). For the same initial perturbation, the dynamical path of successive bundle aggregation depends on the control parameter $V$. As examples, we see that for $V/d=0.92$, quadrimers sweep through the domain right after the dimers form and the system reaches the stable equilibrium, for $V/d=0.80$ the dimers persist for a while before they eventually collapse to quadrimers, and for $V/d=0.70$ the dimers stay stable, as shown in Figure \ref{differentVs}(c).

Having seen how the system coarsens when the volume is controlled, we now turn to the more realistic case when the volume itself evolves dynamically.

\subsubsection{Coupling to dynamics of drying}\label{subsubsec:2ddrying}
For an evaporation dominated situation, the rate at which the liquid volume in each cell is reduced depends on the local surface area of the air-liquid interface and thus on deflection angles of the adjacent lamellae. Consequently, the drying dynamics is coupled to the evolution of the geometric configuration, and new instabilities associated with inhomogeneous cell volumes are expected, which cannot be captured by either energy minimization \cite{Bico2004, Chandra2009, Zhao2006} or renormalization analysis \cite{Gat2013}.

\begin{figure}
  \centering
  \includegraphics[width=1\textwidth]{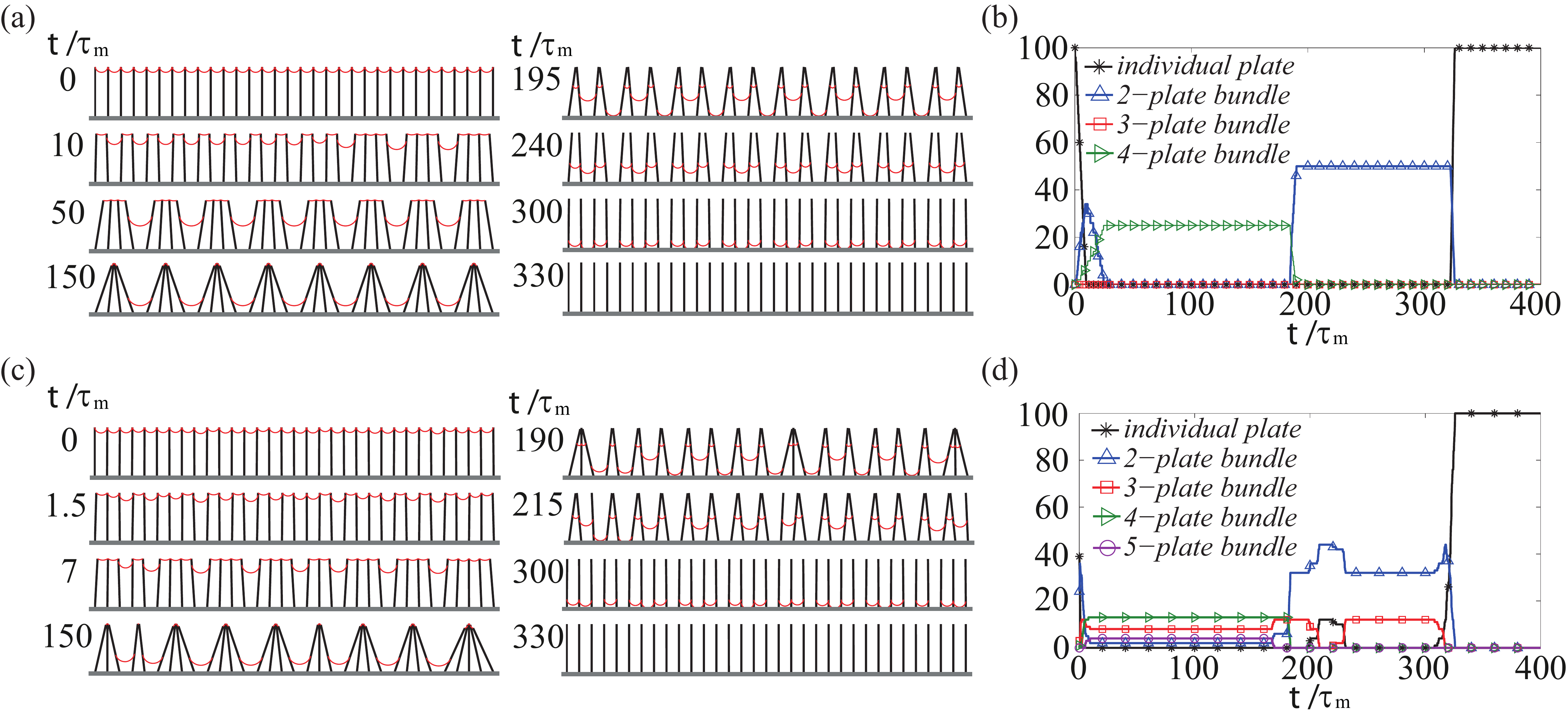}
  \caption{Drying induced elastocapillary coalescence of  plates assuming evaporation rates to depend on the surface area in each cell. {A periodic domain of 100 plates is simulated by solving the coupled Eqs. (\ref{Eq:1storder}) and (\ref{Eq:dvdt}), of which only 32 are displayed.}  (a) Snapshots of the array of plates and menisci at different times. The initial conditions are $V_n/d=0.92$ except that $V_{99}$ is smaller by $2\%$ and $\theta_n=90^o$. (b) Number count of bundles of different size as a function of time for (a). (c) Same as (a) but with uniform random initial perturbations with the maximum amplitude of $5\%$ on both $V_n/d=0.92$ and $\theta_n=90^o$. (d) Number count of bundles of different size as a function of time for (c). Note the 3- and 5-plate bundles.}
  \label{evaporation}
\end{figure}

A minimal evaporation model that is sufficient to capture the qualitative features of hierarchical bundle formation is given by
\begin{equation}
\frac{d V_n}{d t} = \left\{ \begin{array}{l}  2 \frac{t_m}{t_e} c_e r_n\beta_n \quad \text{when \quad $V_n>0$},\\
0 \quad \text{otherwise},
\end{array}
\right.
\label{Eq:dvdt}
\end{equation}
where $c_e$ is a constant,  $\tau_m$ and $\tau_e$ are time scales for mechanical relaxation and evaporation respectively,  $r_n$ is the radius of the meniscus and $2\beta_n$ is the angle subtended by the meniscus arc. Eq. (\ref{Eq:1storder}) and Eq. (\ref{Eq:dvdt}) coupled together determine the dynamics of lamella coalescence driven by evaporating liquid, with three regimes. When $t_m\ll t_e$, $V_n$ decreases quasi-statically so that Eq. (\ref{Eq:critical_dynamics}) relaxes to a static state for a prescribed value of $V_n$. When $t_m\sim t_e$, the evolution of the cell volumes $V_n$ and the lamellar configuration $\theta_n$ are coupled. When $t_m\gg t_e$, the evaporation is so fast that the lamella array does not coarsen. In our typical experiments, $\tau_m\sim 22.5 ms$ and $\tau_e$ is of the order of seconds, so we choose $c_e$ correspondingly in Eq. (\ref{Eq:dvdt}) so that the liquid dries out in about $7$ seconds.  Figure \ref{evaporation} shows the simulation results for a periodic domain of 100 plates with 2 different initial perturbations.

In Figure \ref{evaporation}(a), all plates are perfectly vertical, and the liquid volume in each cell is constant for all cells except that $V_{99}$ is smaller by $2\%$ to mimic boundary effects in an experimental system. The evolution of the system indicates  that a front of dimer coalescence propagates from the imperfection site and sweeps through the entire domain, followed by a successive front of quadrimer coalescence. The largest transient bundles have 4 lamellae. If adhesion between contacting lamella is neglected, capillary forces are not sufficiently large to hold the plates together, and the bundles separate symmetrically when the liquid volume falls bellow a second threshold, and a perfectly vertical configuration is restored. Figure \ref{evaporation}(b) shows the number of bundles of different sizes as a function of time corresponding to the configuration shown in Figure \ref{evaporation}(a), and highlights the fact that bundle formation/separation is perfectly hierarchical and regular.

In Figure \ref{evaporation}(c), uniform random perturbations with maximum relative amplitude of $5\%$ are applied to all the tilting angles and liquid volume in all cells. Irregular bundles of size ranging from 2 to 5 form transiently, and separate as the liquid evaporates. Figure \ref{evaporation}(d) presents the number count of bundles of different size for Figure \ref{evaporation}(c), which shows that although the dimer is still the dominant mode in the early stages, bundle formation can be irregular. This is because the the jump in the location of the contact line when the contact angle reaches a critical value leads to a sudden decrease in the effective stiffness as shown in Figure \ref{super}(a), and the uniform random initial perturbation generates multiple sites from which the front of dimer coalescence starts propagating. Consequently the sites are not necessarily separated by an even number of plates. Therefore, trimers and pentamers also arise in addition to quadrimers. Bundles of larger size do not appear because of the large elastic energy associated with their formation.

The dynamics of coalescence in these two cases shows how the number of plates per cluster varies in a step-like manner, very similar to the experimental data reported by Pokroy \cite{Pokroy2009} and Gat \cite{Gat2013}. All these features also agree qualitatively well with our own experimental observations shown in Figure \ref{experiments}(a)(b).

\section{Collective dynamics of a two-dimensional array of pillars}\label{sec:3d}

\subsection{Experimental observations}\label{subsec:3dexperiments}
We now generalize our study of the one dimensional dynamics of plate or lamella aggregation driven by capillarity to  the coalescence of a two-dimensional array of epoxy nano-pillars immersed in an evaporating wetting liquid as reported in detail in previous work \cite{ Pokroy2009, Kang2010}. The dynamics of pillars is different from that of plates in 3 major ways. First, fluid can flow freely around the multi-connected domains associated with pillars, so that the interaction between them occurs over much longer ranges rather than being limited to just nearest neighbors.  Secondly, the three-dimensional geometry allows the pillars to bend in two principal directions and also twist. For pillars with a circular cross-section, the twist must be a constant. If  we neglect  friction between adhering filaments, the twist must identically vanish, and here we will assume that this is the case.   Thirdly, experimentally we see that a segmented, ``wormlike" geometry of the specially treated pillars increases pinning of the receding contact line by reentrant curvature \cite{ Pokroy2009}; here will neglect this effect for simplicity.

As in the case of lamellae, the uniform array of non-interacting straight pillars loses stability as the liquid evaporates. The dynamics of the ensuing structures is a result of the competition between elasticity and capillarity, and the morphology of the final assembly is determined by intrapillar elasticity and interpillar adhesion \cite{Kang2010}. Figure \ref{experiments}(c)-(e) show the scanning electron microscopy (SEM) images of the assembly into uniform periodic fourfold clusters of nanopillars, in which the pillar height $L=4.5\mu m$, the pillar radius $R=150 nm$, the pillar spacing $D=2\mu m$, the Young's modulus $E=0.2GPa$, the surface tension of the liquid $\sigma =0.022 N/m$, and the density of the liquid $\rho = 786Kg/m^3$. Unlike in the one-dimensional array of lamellae, where the dimer is the primary unstable mode, for pillars the quadrimer is the primary unstable mode. As the liquid evaporates, this mode gives way to hierarchically grow into larger assemblies which eventually get arrested by the increase in the elastic resistance.

\subsection{Mechanics: coupling filament deformation to fluid interface shape}\label{subsec:3dmodel}

\begin{figure}
  \centering
  \includegraphics[width=0.75\textwidth]{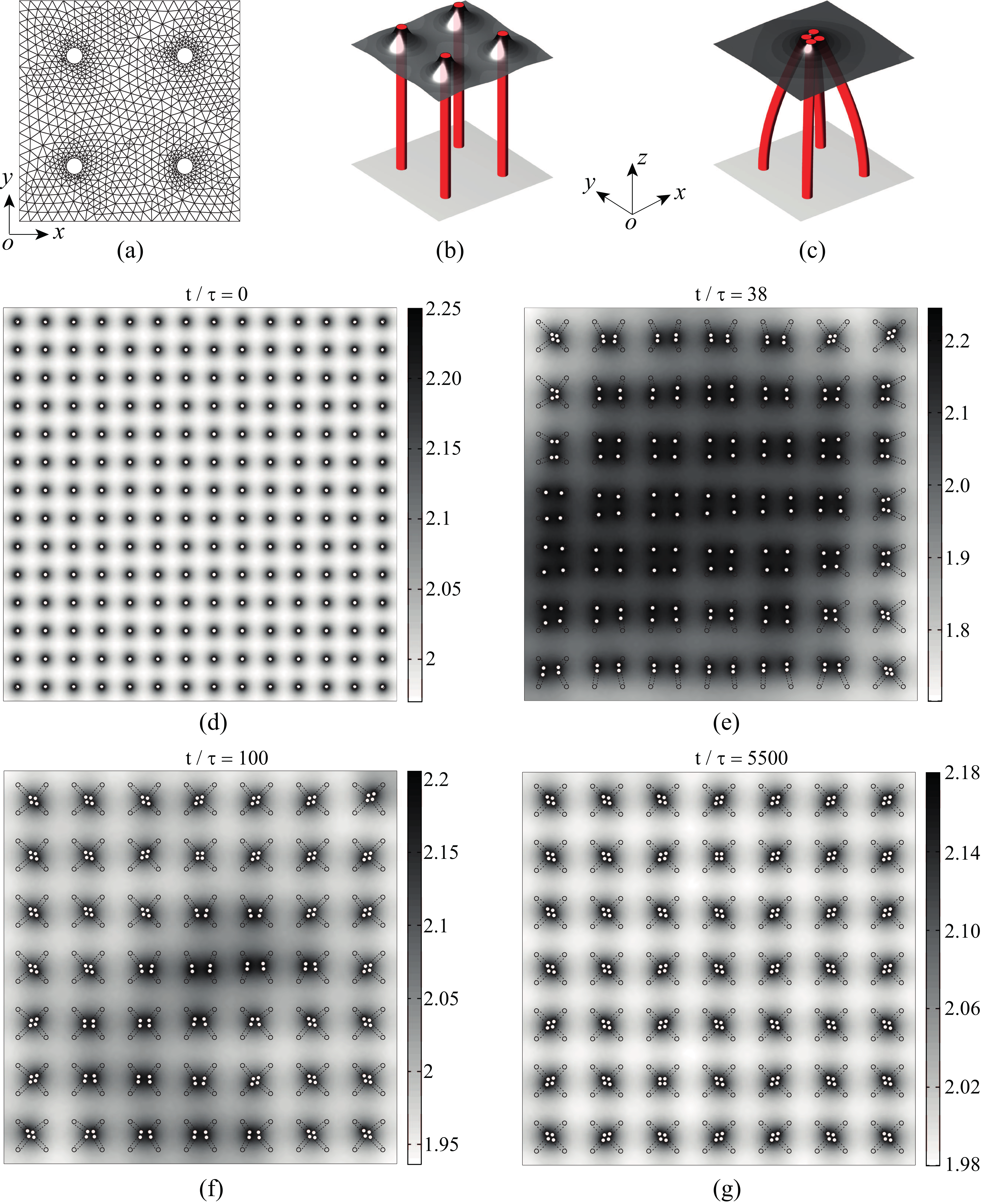}
  \caption{Elastocapillary coalescence of a square array of pillars. (a) Demonstration of triangular meshes on a domain that contains an array of 2 by 2 vertical pillars, where the white solid circles represents pillar tips. The mesh density used in the actual simulation is 4 times denser. (b) The three-dimensional air-liquid interface is obtained by solving Eq. (\ref{Eq:2Dinterface}) on the domain shown in (a) for a given liquid volume $V/V_{flat}=0.85$ in Eq. (\ref{Eq:2DliquidVol}). (c) Given a slight perturbation to the vertical state in (b), 4 pillars coalesce to form a bundle. (d)-(g) A domain of 14 by 14 pillars evolves to the steady state of fourfold clusters for a given liquid volume $V/V_{flat}=0.90$. The dynamics follows the coupled evolution equations (\ref{Eq:2ddynamics})-(\ref{Eq:2DliquidVol}). $\tau$ is the dimensionless time scaled by $\tau_m$ (see text). The gray scale shows the air-liquid interface height, scaled by the pillar spacing. White solid circles represent pillar tips, and the black open circles represent pillar bases. The dashed lines connecting bases and tips correspond to a projection of pillars viewed from the top. }
  \label{2dsim}
\end{figure}

As in the lamellar case, inertial effects can be neglected here, so that the dynamics of each pillar tip can be characterized by its displacement vector relative to its base $\overrightarrow{X}(x,y)$ (Figure \ref{2dsim}(a)), and is given by
\begin{equation}
c\frac{d\overrightarrow{X}}{dt} + \overrightarrow{F}_b(\overrightarrow{X}) + \overrightarrow{F}_{\sigma}(\overrightarrow{X},V)=0,
\label{Eq:2ddynamics}
\end{equation}
where $c$ is the drag coefficient, $\overrightarrow{F}_b$ is the elastic bending resistance force at the tip, $\overrightarrow{F}_{\sigma}$ is the capillary driving force due to surface tension $\sigma$, and $V$ is the liquid volume in the system. The dominant contribution to $c$ is from the internal damping of the viscoelastic solid similar to the lamellar case, and $c\approx 3 t_m\pi E R^4/4 L^3$ {(APPENDIX \ref{AppendixD})}, where $t_m\sim10^{-2}s$ is the time scale for the fiber to relax mechanically.

To compute the bending resistance force $\overrightarrow{F}_b$ in the horizontal direction, we use the theory of the elastica for the inextensional, unshearable deformation of thin filament. Letting $\vartheta$ be the angle of the pillar centerline tangent with the vertical direction, with $s\in[0,L]$ is the arc length coordinate, and $|\overrightarrow{F}_b|$  the  force amplitude at the tip, equilibrium implies that
\begin{equation}
\frac{\pi}{4} R^2 E \vartheta_{ss} + |\overrightarrow{F}_b| \cos\vartheta=0,
\label{Eq:2Drod}
\end{equation}
Geometry implies that $|\overrightarrow{X}| = \int_0^L \cos[\vartheta(s;|\overrightarrow{F}_b|)]ds$, so given $\overrightarrow{X}$, $|\overrightarrow{F}_b|$ is uniquely determined, and $\overrightarrow{F}_b = |\overrightarrow{F}_b|\overrightarrow{X}/|\overrightarrow{X}|$.

To obtain $\overrightarrow{F}_{\sigma}$, we need to determine the shape of the air-liquid interface.
Since the Bond number $Bo\sim10^{-6}\ll1$, gravity can be neglected. Moreover, the time scale for the fluid to equilibrate in the porous brush $t_f$ is much smaller than that for the pillars to relax mechanically $t_m$, which is much smaller than that for the evaporation $t_e$, i.e. $t_f\sim10^{-3}s \ll t_m\sim10^{-2}s\ll t_e\sim 10^0s$ {(APPENDIX \ref{AppendixE})}. Therefore, the pressure throughout the liquid domain can be regarded as uniform and the air-liquid interface $z=S(x,y)$ is thus a surface of  uniform mean curvature and satisfies the equation
\begin{equation}
2 \sigma H = \sigma \frac{(1+S_x^2)S_{yy}-2S_xS_yS_{xy}+(1+S_y^2)S_{xx}}{(1+S_x^2+S_y^2)^{3/2}}=  p,
\label{Eq:2Dinterface}
\end{equation}
where $H$ is the mean curvature of the interfacial surface. Without loss of generality, we set the ambient pressure to zero, and let $p$ be the pressure inside the liquid. Volume conservation in the whole domain yields
\begin{equation}
V=\int_A S(x,y)dxdy + \pi R^2\displaystyle\sum_{i=1}^{N} (h_i-L),
\label{Eq:2DliquidVol}
\end{equation}
and serves to determine $p$. Here $A$ is the projected domain of the air-liquid interface to the horizontal plane (meshed area in Figure \ref{2dsim}(a)), and $h_i$ is the elevation of $i^{th}$ pillar tip. Since the menisci are always pinned on the pillar tips, we need to solve Eq. (\ref{Eq:2Dinterface}) on a multiply connected domain, in which pillar tips are regarded as solid circles with the identical radius (Figure \ref{2dsim}(a)) and the height of the surface is fixed at the elevation of pillar tips (Figure \ref{2dsim}(b)) $h=\int_0^L \sin\vartheta ds$ calculated from Eq. (\ref{Eq:2Drod}). As the pillars are effectively immersed in liquid, the integration of pressure over the lateral surface of the cylinder does not contribute to $\overrightarrow{F}_{\sigma}$, and the only active contribution is the line tension at the contact line. For a given air-liquid interface (Figure \ref{2dsim}(c)), the angle of the meniscus tangent with the horizontal direction on the circular boundary of the tip is known, which we denote as $\varphi$. To calculate the capillary driving force on each pillar, we integrate the interfacial force over the contact line contour at the tip and determine the component in the horizontal direction that contributes to the deflection, with $\overrightarrow{F}_{\sigma}=\oint_c \sigma \overrightarrow{n} \cos\varphi ds$, where the subscript $c$ represents the tip circle and $\overrightarrow{n}$ is { its unit outward normal}. Note that for very small values of $V$, the assumption of the surface being pinned at the pillar tips breaks down; however clusters form well before this assumption is violated, so that this is not a cause for concern.

To prevent penetration upon collision between pillars, we treat each pillar as a rod with finite radius with an artificial short-range repulsion force when the discs representing pillar tips come close enough ($10\%$ of the pillar diameter), but do not consider the elastic deformation of the cross section due to contact.

\begin{figure}
  \centering
  \includegraphics[width=1\textwidth]{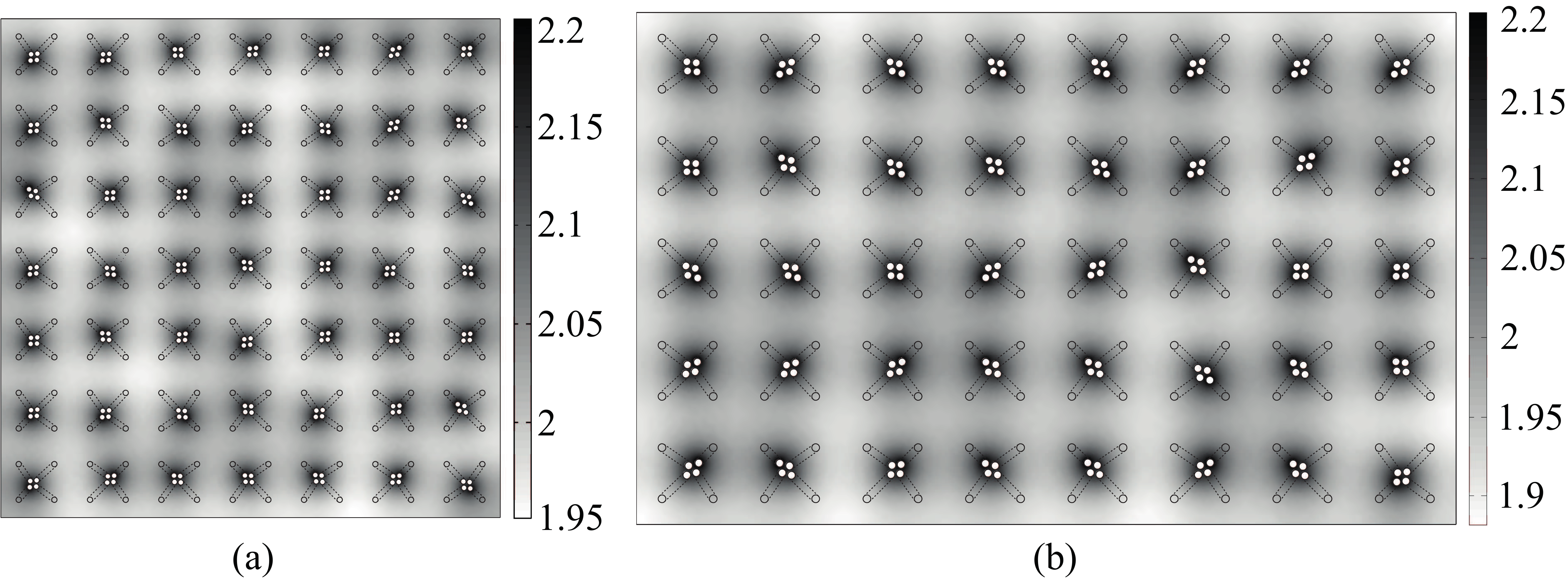}
  \caption{More examples of final structures of the fourfold clusters for different prescribed volumes and domain sizes with random initial perturbations and symmetric boundary conditions. (a) 14 by 14 pillars with a given liquid volume $V/V_{flat}=0.90$.  (b) 10 by 16 pillars with a given liquid volume $V/V_{flat}=0.88$. Simulations were done by solving equations (\ref{Eq:2ddynamics})-(\ref{Eq:2DliquidVol}).}
  \label{othersamples}
\end{figure}

\subsection{Nonlinear dynamics: simulations  and comparison with experiments}\label{subsec:3dsimulation}

To complete the formulation of the problem, we need some boundary conditions. We consider a square array of pillars inside a domain with straight vertical walls and impose symmetry-related conditions on the walls for the fluid interface and the pillar deformations. On the pillars, the contact lines are assumed to be pinned on all pillar tips.  For initial conditions, the pillar bases (open circles in Figure \ref{2dsim}(g)) are assumed to form a perfect periodic square lattice, while the pillar tips (solid white circles in Figure \ref{2dsim}(g)) are perturbed from the vertical configuration so that the layer of pillars closest to the boundaries inclines inwards to trigger the inward motion from the boundaries, and other pillar tips are uniformly and randomly perturbed. We numerically solve the coupled evolution equations (3.1)-(3.4) using a custom-coded finite element code. For a given pillar tip displacement $\overrightarrow{X_i}$, Eq. (\ref{Eq:2Drod}) is solved with the geometric condition $|\overrightarrow{X_i}| = \int_0^L \cos[\vartheta_i(s;|\overrightarrow{F}_{bi}|)] ds$ to determine the reaction force $\overrightarrow{F}_{bi}$ and tip elevation of each pillar $h_i$. Given $\overrightarrow{X_i}$ and $h_i$, Eq. (\ref{Eq:2Dinterface}) is solved using the finite element method to determine the interface $S(x,y)$ on the multiply-connected domain with the liquid volume constraint Eq. (\ref{Eq:2DliquidVol}), which determines $p$ and thence $\overrightarrow{F}_{\sigma i}$. Then Eq. (\ref{Eq:2ddynamics}) is integrated in time explicitly to update the pillar tip positions, and the domain is remeshed accordingly at every time step.

Figure \ref{2dsim}(d)-(g) show the simulated evolution of an array of $14$ by $14$ pillars collapsing into fourfold bundles for a prescribed liquid volume of $V/V_{flat}=0.90$, where $V_{flat}$ is the control volume inside the system when all the pillars are vertical and the air-liquid interface is flat, using the parameters from experiments in Figure \ref{experiments}(c)-(e). Due to the initial boundary perturbations, coalescence is initiated from the boundary and propagates towards the center of the domain. We observe that there is a critical liquid volume above which uniform vertical pillars are stable, and below which pillar coalescence occurs, similar to the case of lamellar collapse. However, here the primary eigenmode of instability has a fourfold symmetry (Figure \ref{2dsim}(e))  independent of initial perturbations as observed in experiments (Figure \ref{2dsim}(g) and Figure \ref{othersamples}). An intuitive way to understand this is to recognize that the fourfold bundles have two principal directions, along which one sees dimers. Although the interaction potential between pillars in the three-dimensional case has a much longer range than in the two-dimensional case because of connectivity, it is still monotonic and decreases with distance. Provided that within the linear analysis the effective spring constants in the two principal directions are decoupled, the dimer is the primary mode in each direction as in the one-dimensional lamella array. Beyond the linear regime, our simulations capture the coarsening that is arrested and eventually leads to a maximum bundle size.  We also note that the tips in a bundle can form either rhombi or squares depending on the initial perturbations, although rhombi are more likely as they are stable against shear deformations; indeed as we neglect friction between the tips, the rhombus is a more energetically favorable configuration than the square. However, the energy difference between these two states is very small, so that contact and friction in real experiments leads to both square tips (Figure \ref{experiments}(d)) and rhombi tips (Figure \ref{experiments}(c)).

\section{Conclusions}\label{sec:conclusion}

Our study has focused on understanding the onset and evolution of elasto-capillary coalescence of plates and pillars driven by evaporation. For the case of lamellar collapse, we explicitly derived the conditions for the primary dimerization instability in terms of the state variables - tilting angles $\theta_n$ and liquid volumes $V_n$, and the relevant geometrical and physical parameters. Complementing our analysis, full numerical simulations show that the final coalescent states sensitively depend on initial perturbations because of the discontinuous motion of the contact line when the contact angle reaches a critical value.  This implies that the self-organization of clusters cannot be predicted by energy minimization arguments alone, but depend on the dynamics of the drying process - this is especially true when the coupling of geometry to local evaporation rates is taken into account. Our model accounts for this, and is in qualitative agreement with experimental observations of the intermittence of coalescent transitions. For the case of pillar collapse, our model correctly accounts for the multiply-connected nature of the fluid interface, and the large elastic deflections of the pillars. The analysis based on this model captures the primary fourfold eigenmode associated with the onset of collapse, consistent with experimental observations. Numerical simulations of the  full dynamics allow us to follow the evolution of the clusters whose eventual size is determined by the competition between capillarity and elasticity. For both cases, our numerical results agree well with many of the salient experimental observations. In particular, we can explain the eigenmodes at the onset of instability, and the time scales on which clusters form, while providing explanations for  both regular and irregular hierarchical bundling till the final state.

However, our analysis still leaves out some effects and thus cannot explain some observations.  Neglecting adhesion and friction between pillars implies that we cannot explain the twisting of pillars that leads to the formation of chiral clusters often seen. This is a natural next step in the analysis. Furthermore, we have limited ourselves to a discrete theory in both cases, but in the thermodynamic limit of a large number of pillars or lamellae, one might ask what the nature of a continuum theory might be.  A recent continuum theory that addresses the explicit connection of the essential geometric and physical parameters to determine the maximal size and dynamics of the assembly has been carried out for lamellar coalescence \cite{Wei2014}, but the question for 3-dimensional coalescence remains an open question.

\section*{Acknowledgment}
We thank Sam Ocko for many discussions that helped to sharpen and clarify our arguments. We thank the Harvard-MRSEC DMR -0820484, the MacArthur Foundation (LM), { and the NRF of Korea (Grant No. 2013034978, H.-Y.K.)} for support.

\bibliographystyle{prs}

\appendix
\section{Critical buckling length of a plate}\label{AppendixA}
The critical buckling height of a thin plate under compression due to  surface tension $\sigma$ is
\begin{equation}
L_{crit}\sim L_{ec} =  \sqrt{Eh^3/\sigma},
\end{equation}
where $L_{ec}$ is the typical elastocapillary length scale. For the case when the plate is clamped at one end and axially compressed at the other, the exact expression \cite{Timoshenko1961} for $L_{crit}$ is
\begin{equation}
L_{crit} = \pi [48(1-\nu^2)]^{-1/2} L_{ec} \approx 1mm,
\end{equation}
where we have substituted in the experimentally observed parameter values. As $L_{crit}$ is much larger than the lamella height $L=40\mu m$, buckling of the lamellae when they pierce the gas-liquid interface does not happen in our system.

\section{Moment calculation for the 2D case}\label{AppendixB}
The 8 possible meniscus configurations can be classified into 6 cases, where we explicitly express the moments in terms of the state variables - tilting angles $\theta_n$ and liquid volumes $V_n$. We have scaled all lengths by $L$ and moment by $\sigma L$, so all results below are dimensionless.

{\bf Case 1:} The meniscus is pinned on both tips (Figure \ref{model}(b1)). The half angle $\beta_n$ subtended by the meniscus arc is determined by solving
    \begin{equation}
    V_n = \frac{1}{2}d(\sin\theta_n + \sin\theta_{n+1})+\frac{1}{2}\sin(\theta_n-\theta_{n+1})-a_n^2(\beta_n\csc^2\beta_n - \cot\beta_n),
    \label{Eq_S:pin_V}
    \end{equation}
    for given $V_n$, $\theta_n$ and $\theta_{n+1}$, where $d=D/L$ and half of the tip distance $a_n$ is
    \begin{equation}
    a_n = \frac{1}{2}\sqrt{2-2\cos(\theta_{n+1}-\theta_n)+2d(\cos\theta_{n+1}-\cos\theta_n)+d^2}.
    \label{Eq_S:pin_a}
    \end{equation}
    $\beta_n$ must satisfy $\beta_n\leq\hat{\beta}_n$, where $\hat{\beta}_n$ is the critical angle at which the meniscus starts to slide down from at least one lamella. $\beta_n<0$ when the meniscus concaves down, $\beta_n=0$ when the meniscus is flat, and $\beta_n>0$ when the meniscus concaves up. The moments on the $n^{th}$ and $(n+1)^{th}$ plates are given respectively by
    \begin{align}
    \label{Eq_S:pin_M1}
    &M_n = -\frac{1}{2a_n}\left[\sin(\beta_n+\theta_n-\theta_{n+1})+d\sin(\beta_n+\theta_n)\right],\\
    \label{Eq_S:pin_M2}
    &M_{n+1} = \frac{1}{2a_n}\left[\sin(\beta_n+\theta_n-\theta_{n+1})-d\sin(\beta_n-\theta_{n+1})\right],
    \end{align}

{\bf Case 2:} The meniscus is down from both tips (Figure \ref{model}(b3)). The contact angle is fixed at $\alpha$. When $\theta_n\neq\theta_{n+1}$, the radius of the meniscus $r_n$ is determined by solving
\begin{equation}
\begin{split}
V_n = &\frac{d^2}{2}\frac{\sin\theta_n\sin\theta_{n+1}}{\sin(\theta_{n+1}-\theta_n)}-r_n^2\left[\cos^2\alpha \cot\left(\frac{\theta_{n+1}-\theta_n}{2}\right)\right.\\
    & \left.+ \sin\alpha\cos\alpha +\frac{\pi+\theta_{n+1}-\theta_n-2\alpha}{2}\right].
\end{split}
\label{Eq_S:bothdown_V}
\end{equation}
The wetting length on the right side of the $n^{th}$ plate and that on the left side of the $(n+1)^{th}$ are given respectively  by
\begin{align}
\label{Eq_S:bothdown_Lwr}
{L_{wr}}_n &= d\frac{\sin\theta_{n+1}}{\sin(\theta_{n+1}-\theta_n)}-r_n\left[\sin\alpha+\cos\alpha\cot\left(\frac{\theta_{n+1}-\theta_n}{2}\right)\right].\\
\label{Eq_S:bothdown_Lwl}
{L_{wl}}_{n+1} &= d\frac{\sin\theta_n}{\sin(\theta_{n+1}-\theta_n)}-r_n\left[\sin\alpha+\cos\alpha\cot\left(\frac{\theta_{n+1}-\theta_n}{2}\right)\right].
\end{align}

When $\theta_n=\theta_{n+1}=\theta$, the meniscus radius is independent of $V_n$, and $r_n=d\sin\theta/(2\cos\alpha)$. $l_n$ is determined by solving
\begin{equation}
V_n = \frac{2l_n-d\cos\theta}{2}d\sin\theta - \left(\frac{d\sin\theta}{2}\right)^2\tan\alpha - \frac{\pi-2\alpha}{2}\left(\frac{d\sin\theta}{2\cos\alpha}\right)^2.
\label{Eq_S:bothdown_V2}
\end{equation}
The wetting length on the right side of the $n^{th}$ plate and that on the left side of the $(n+1)^{th}$ plate are given respectively by
\begin{align}
\label{Eq_S:bothdown_Lwr2}
{L_{wr}}_n &= l_n-\frac{d}{2}\sin\theta\tan\alpha,\\
\label{Eq_S:bothdown_Lwl2}
{L_{wl}}_{n+1} &= l_n-\frac{d}{2}\sin\theta\tan\alpha-d\cos\theta.
\end{align}

In either case, the moment on the $n^{th}$ plate and that on the $(n+1)^{th}$ plate are given respectively by
\begin{align}
\label{Eq_S:bothdown_M1}
M_n &= -\frac{{L_{wr}}_n^2}{2r_n}-{L_{wr}}_n\sin\alpha.\\
\label{Eq_S:bothdown_M2}
M_{n+1} &= \frac{{L_{wl}}_{n+1}^2}{2r_n}+{L_{wl}}_{n+1}\sin\alpha.
\end{align}

{\bf Case 3:} when the meniscus slides down from the $n^{th}$ plate and is pinned on the $(n+1)^{th}$ one. In this case ${L_{wr}}_n=l_n$ and ${L_{wl}}_{n+1}=1$.
\begin{equation}
a_n = \frac{1}{2}\sqrt{1+l_n^2-2l_n\cos(\theta_n-\theta_{n+1})+2d(\cos\theta_{n+1}-l_n\cos\theta_n)+d^2},
\end{equation}
\begin{equation}
V_n = \frac{1}{2}l_n\sin(\theta_n-\theta_{n+1})+\frac{1}{2}d(l_n\sin\theta_n+\sin\theta_{n+1})-a_n^2(\beta_n\csc^2\beta_n-\cot\beta_n).
\label{Eq_S:leftdown_V}
\end{equation}

From the condition that contact angle on the $n^{th}$ plate is $\alpha$, we can get the following relations,
\begin{equation}
\cos(\alpha+\theta_n-\theta_{n+1})+d\cos(\alpha+\theta_n)-l_n\cos\alpha = 2a_n\cos\beta_n,
\label{Eq_S:leftdown_lbeta1}
\end{equation}
and
\begin{equation}
\begin{array}{l}
\sin(\alpha+\beta_n+\theta_n-\theta_{n+1})+d\sin(\alpha+\beta_n+\theta_n) = l_n\sin(\alpha+\beta_n)\\
 \text{when} \quad \theta_n \neq \theta_{n+1},\\
d\cos\theta + 1-l_n = \cos(\alpha+\beta_n) \\
 \text{when} \quad \theta_n=\theta_{n+1}=\theta.
\end{array}
\label{Eq_S:leftdown_lbeta2}
\end{equation}
$l_n$ and $\beta_n$ are determined by solving either (\ref{Eq_S:leftdown_V}) and (\ref{Eq_S:leftdown_lbeta1}) or (\ref{Eq_S:leftdown_V}) and (\ref{Eq_S:leftdown_lbeta2}).

Moment on the $n^{th}$ plate and that on the $(n+1)^{th}$ plate are given respectively by
\begin{align}
\label{Eq_S:leftdown_M1}
M_n &= -\frac{\sin\beta_n}{2a_n}l_n^2-l_n\sin\alpha.\\
\label{Eq_S:leftdown_M2}
M_{n+1} &= \frac{\sin\beta_n}{2a_n}-\sin(\theta_{n+1}-\theta_n-\alpha-2\beta_n).
\end{align}

{\bf Case 4:} when the meniscus slides down from the $(n+1)^{th}$ plate and is pinned on the $n^{th}$ one. In this case ${L_{wr}}_n=1$ and ${L_{wl}}_{n+1}=l_n$.

\begin{equation}
a_n = \frac{1}{2}\sqrt{1+l_n^2-2l_n\cos(\theta_n-\theta_{n+1})+2d(l_n\cos\theta_{n+1}-\cos\theta_n)+d^2},
\label{Eq_S:rightdown_a}
\end{equation}
\begin{equation}
V_n = \frac{1}{2}l_n\sin(\theta_n-\theta_{n+1})+\frac{1}{2}d(\sin\theta_n+l_n\sin\theta_{n+1})-a_n^2(\beta_n\csc^2\beta_n-\cot\beta_n).
\label{Eq_S:rightdown_V}
\end{equation}

From the condition that contact angle on the $(n+1)^{th}$ plate is $\alpha$, we can get the following relations,
\begin{equation}
\cos(\alpha+\theta_n-\theta_{n+1})-d\cos(\alpha-\theta_{n+1})-l_n\cos\alpha = -2a_n\cos\beta_n,
\label{Eq_S:rightdown_lbeta1}
\end{equation}
and
\begin{equation}
\begin{array}{l}
\sin(\alpha+\beta_n+\theta_n-\theta_{n+1})-d\sin(\alpha+\beta_n-\theta_{n+1}) = l_n\sin(\alpha+\beta_n) \\
\text{when} \quad \theta_n \neq \theta_{n+1},\\
d\cos\theta + l_n-1 = \cos(\alpha+\beta_n) \\
\text{when} \quad \theta_n=\theta_{n+1}=\theta.
\end{array}
\label{Eq_S:rightdown_lbeta2}
\end{equation}
Similarly, $l_n$ and $\beta_n$ are determined by solving either (\ref{Eq_S:rightdown_V}) and (\ref{Eq_S:rightdown_lbeta1}) or (\ref{Eq_S:rightdown_V}) and (\ref{Eq_S:rightdown_lbeta2}).

Moment on the $n^{th}$ plate and that on the $(n+1)^{th}$ plate are given respectively by
\begin{align}
\label{Eq_S:rightdown_M1}
M_n &= -\frac{\sin\beta_n}{2a_n}+\sin(\theta_{n+1}-\theta_n-\alpha-2\beta_n).\\
\label{Eq_S:rightdown_M2}
M_{n+1} &= \frac{\sin\beta_n}{2a_n}l_n^2+l_n\sin\alpha.
\end{align}

For certain given $V_n$, $\theta_n$ and $\theta_{n+1}$, although one end of the meniscus is depinned from the tip, there is no arc satisfying the enforced contact angle condition. In the following two cases, the menisci are approximated by straight lines. They are used to prevent the numerics from blowing up when two plates almost contact, yet very unlikely to happen.

{\bf Case 5:} when the meniscus slides down from the tip of the $n^{th}$  plate and keeps flat. The moment on the $n^{th}$ plate and that on the $(n+1)^{th}$ plate are given respectively by
\begin{align}
M_n &= -\frac{l_n}{b_n}[\sin(\theta_n-\theta_{n+1})+d\sin\theta_n]\\
M_{n+1} &= \frac{1}{b_n}[l_n\sin(\theta_n-\theta_{n+1})+d\sin\theta_{n+1}],
\end{align}
where $l_n$ is determined by solving from the given volume
\begin{equation}
V_n = \frac{l_n}{2}\sin(\theta_n-\theta_{n+1})+\frac{d}{2}(l_n\sin\theta_n+\sin\theta_{n+1}),
\end{equation}
and $b_n$ is
\begin{equation}
b_n=\sqrt{1+l_n^2-2l_n\cos(\theta_n-\theta_{n+1})+2d(\cos\theta_{n+1}-l_n\cos\theta_n)+d^2}.
\end{equation}

{\bf Case 6:} when the meniscus slides down from the tip of the $n^{th}$ plate and keeps flat. The moment on the $n^{th}$ plate and that on the $(n+1)^{th}$ plate are given respectively by
\begin{align}
M_n &= -\frac{1}{b_n}[l_n\sin(\theta_n-\theta_{n+1})+d\sin\theta_n]\\
M_{n+1} &= \frac{l_n}{b_n}[\sin(\theta_n-\theta_{n+1})+d\sin\theta_{n+1}],
\end{align}
where $l_n$ is determined by solving from the given volume
\begin{equation}
V_n = \frac{l_n}{2}\sin(\theta_n-\theta_{n+1})+\frac{d}{2}(\sin\theta_n+l_n\sin\theta_{n+1}),
\end{equation}
and $b_n$ is
\begin{equation}
b_n=\sqrt{1+l_n^2-2l_n\cos(\theta_n-\theta_{n+1})+2d(l_n\cos\theta_{n+1}-\cos\theta_n)+d^2}.
\end{equation}

\section{Damping coefficient for the 2D case of plates}\label{AppendixC}
\begin{figure}
  \centering
  \includegraphics[width=0.7\textwidth]{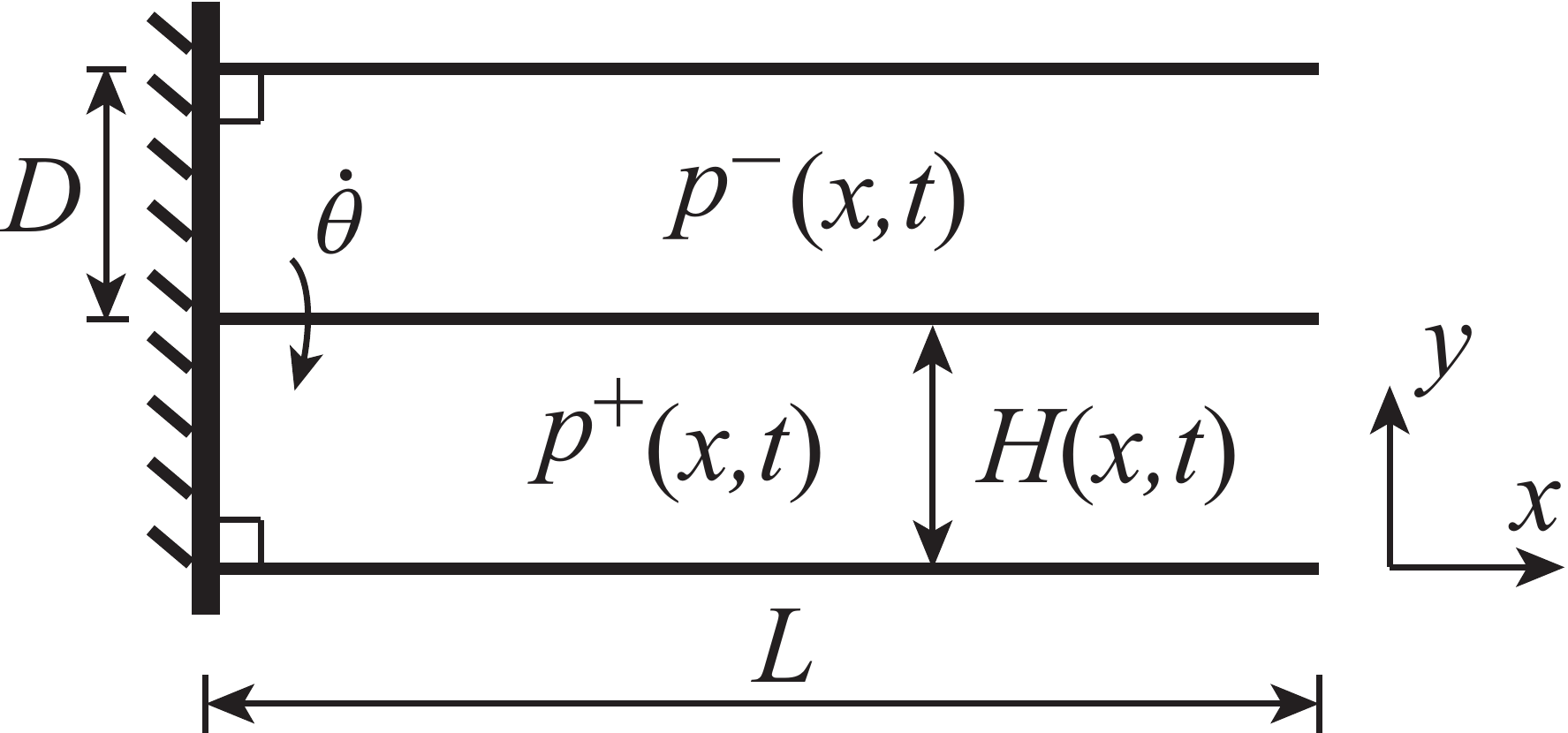}
  \caption{Sketch of two neighboring cells. When the middle plate is rotating, the nonuniform pressure caused by flow acts on the plate from both sides and hinders the rotation.}
  \label{lubri}
\end{figure}

In order to calculate the damping coefficient, we need to account for both the viscosity of the fluid and the viscoelasticity of the solid. First we consider the contribution from the fluid. The Reynolds number $Re\sim10^{-5}$ or less, so that inertia of the fluid is negligible. We use lubrication theory to calculate the moment acting on the plate caused by flow although the ratio of plate height $L$ to the spacing $D$ does not strictly satisfy that $L/D\gg1$. Figure \ref{lubri} illustrates 3 rigid plates hinged at the base. The upper and lower one are perpendicular to the substrate, and the middle one is rotating clockwise at the angular velocity $\dot{\theta}$. The liquid is confined between two plates and the ambient pressure is set to be 0.

In the bottom chamber, the momentum conservation of the fluid in the $x$ and $y$ directions are
\begin{equation}
\frac{\partial p^+}{\partial x} = \mu \frac{\partial^2 u}{\partial y^2} \quad \text{and} \quad \frac{\partial p^+}{\partial y} =0,
\label{lubrigov}
\end{equation}
where $p^+$ is the pressure and $\mu$ is the fluid viscosity. The boundary conditions are
\begin{equation}
u=0 \quad \text{at} \quad y=0 \quad \text{and} \quad y=H(x,t).
\label{lubribc}
\end{equation}
From (\ref{lubrigov}) and (\ref{lubribc}), we can get
\begin{equation}
u(x,y,t) = \frac{1}{2\mu}\frac{\partial p^+}{\partial x}(y^2-Hy).
\label{velocity}
\end{equation}
The mass conservation is
\begin{equation}
\frac{d}{dt}\left[\frac{1}{2}\rho x (D+H)\right] = \rho[Q(0,t)-Q(x,t)],
\label{masscons}
\end{equation}
where $Q(x,t)$ is the flux. As $u=0$ at $y=0$, $Q(0,t)=0$. Substitute (\ref{velocity}) into (\ref{masscons}),
\begin{equation}
\begin{split}
Q(x,t) &= \int_0^H u(x,y,t)dy = \int_0^H \frac{1}{2\mu}\frac{\partial p^+}{\partial x}(y^2-Hy) dy \\
&= -\frac{H^3}{12\mu}\frac{\partial p^+}{\partial x}= -\frac{1}{2}x\frac{dH}{dt}.
\end{split}
\label{Q}
\end{equation}
From (\ref{Q}) and $H(x,t) = D-\dot{\theta} t x$ in the linear regime, we get
\begin{equation}
\frac{\partial p^+}{\partial x} = -\frac{6\mu}{H^3}x^2\dot{\theta}.
\label{pressuregrad}
\end{equation}
Integrate (\ref{pressuregrad}) and use the boundary condition on pressure $p^+(L,t)=0$, and we see that
\begin{equation}
p^+(s,0) = \frac{2\mu L^3}{D^3}\dot{\theta}(1-s^3),
\label{p}
\end{equation}
where $s=x/L\in[0,1]$.

So far we have only considered the pressure acting on the lower side of the middle pillar. The pressure difference between two sides is
\begin{equation}
p^+(s,0)-p^-(s,0) = 2p^+(s,0) = \frac{4\mu L^3}{D^3}\dot{\theta}(1-s^3).
\label{pdiff}
\end{equation}
And the moment on the middle pillar due to fluid flow is
\begin{equation}
M_{\mu}=L^2\int_0^1(p^+(s,0)-p^-(s,0))s ds = \frac{6\mu L^5 \dot{\theta}}{5D^3},
\end{equation}
where the last equality follows from (\ref{pdiff}).

Therefore the damping coefficient from the viscous fluid is approximately
\begin{equation}
C_1 = \frac{M_{\mu}}{\dot{\theta}}=\frac{6\mu L^5 }{5D^3}.
\end{equation}

Next we consider the damping coefficient resulted from viscoelasticity of the solid. We can estimate it from the vibration of a viscoelastic beam made of the Kelvin-Voigt material, which can be represented by a viscous damper and elastic spring connected in parallel. The constitutive law thus is
\begin{equation}
\Sigma = E\epsilon +\eta \dot{\epsilon},
\label{constitutive}
\end{equation}
where $\Sigma$ is the stress, $\epsilon$ is the strain, $E$ is the Young's modulus and $\eta$ is the viscosity. The strain-displacement relation is
\begin{equation}
\epsilon = \frac{y}{r} = -y \frac{\partial^2 u}{\partial x^2},
\label{strain}
\end{equation}
where $r$ is the radius of curvature due to bending, and $u$ is the transverse displacement of the beam. The force balance equation is
\begin{equation}
\frac{\partial^2 M}{\partial x^2} +\rho_s \ddot{u}=0,
\label{force}
\end{equation}
where the moment $M = -\int_A \Sigma y d A$, and $\rho_s$ is the density of the solid material.

Substitute (\ref{strain}) into (\ref{constitutive}), then substitute (\ref{constitutive}) into (\ref{force}), and we can get the vibration equation of a viscoelastic beam, which is
\begin{equation}
E I \frac{\partial^4 u}{\partial x^4} + \eta I \frac{\partial^4 \dot{u}}{\partial x^4} + \rho_s \ddot{u} = 0, \quad \text{where} \quad I = \int_A y^2 d A.
\label{vibration}
\end{equation}
For the thin plate, $E$ is replaced by $E/(1-\nu^2)$, $\rho$ is replaced by $\rho h$ and $I=h^3/12$, where $h$ is the thickness of the plate. We can get two time scales $\tau_{m1} \sim \sqrt{12\rho(1-\nu^2)L^4/Eh^2}$  and $\tau_{m2} \sim \eta(1-\nu^2)/E$ by balancing the first and the third term, the first and the second term of the LHS of (\ref{vibration}) respectively. Substitute the parameters from the experiment $L=40\mu m$, $h=10\mu m$, $E=1.5M pa$, $\nu=0.5$ and $\rho_s = 965 kg/m^3$, $\tau_{m1}\sim 10 \mu s$. We roughly estimate $\eta$ by using the indentation experimental result in the literature \cite{Mata2005}. For the PDMS cured at the base to curing agent ratio of $10:1$, the loss tangent at a vibration rate of $4Hz$ is around $0.09$, which is $[\eta (1-\nu^2) \omega/ E]|_{\omega = 4} = 0.09$, so $\tau_{m2} \approx 22.5 ms$. As $\tau_{m2}\gg\tau_{m1}$, the inertia term is dominated over by the damping term in (\ref{vibration}) for the micro cantilever, and thus can be neglected. Then the damping coefficient due to internal viscosity of the solid is
\begin{equation}
C = C_1+C_2 = \frac{6\mu L^5 }{5D^3} + \tau_{m} k \approx \tau_{m} k.
\end{equation}
where we have omitted the subscript `2' in $\tau_{m2}$ and $\tau_m\approx 22.5 ms$ is the time scale for the viscoelastic fiber to relax mechanically.

\section{Damping coefficient for the 3D case of pillars}\label{AppendixD}

First, To estimate the drag coefficient $c$ and the time scale in Eq. (\ref{Eq:2ddynamics}), we consider both the internal contribution from viscoelasticity of the solid material and the external contribution from the viscous fluid, and find that the former one dominates.

AFM characterization of the epoxy microfibers \cite{Grinthal2011} shows that the force vs displacement curve has a clear hysteresis loop, which reveals that the epoxy nanofibers are viscoelastic although primarily elastic. For a fiber of radius $R=1.5\mu m$, length $L = 9\mu m$, and bending spring constant $k = 3 \pi E R^4/4 L^3 = 17.22N/m$, where $E=1 GPa$ is the Young's modulus, the hysteresis is measured $32\%$ of the total work done at max deflection $x= 1.5\mu m$ and rate of deflection $v= 6 \mu m/s$. Thus, the effective damping coefficient $c_1$ from  the internal viscoelasticity can be calculated from $2 c_1 v x = 0.3\times0.5\times k x^2$, based on the Kelvin-Voigt model.

To calculate the damping coefficient $c_2$ from the external viscous fluid, we consider a circular cylinder of radius $R$ moving with velocity $U$ normal to its axis at small Reynolds number $Re = 2R\rho U/\mu\sim10^{-7}$, in which case, the drag of magnitude $4\pi \mu U/\ln(7.4/Re)$ per unit length was first derived by Lamb \cite{Lamb1911}. As the center of the pillar is moving at a velocity of almost $v/2$ when the cantilever tip is deformed at a velocity of $v$ in the AFM test, we take $c_2 = 2\pi \mu L /\ln(7.4/Re)$ as an approximation.

It turns out that $c_1/c_2\sim 10^7$ for the experimental parameters listed above, which indicates that the external damping from viscous fluid can be neglected and $c\approx c_1$. Therefore, the time scale for the fiber to relax mechanically is $\tau_m = c_1/k\sim 10^{-2} s$ in  Eq. (3.1), and $\tau_m$ is a material property (as in the Prony series) and does not depend on the geometric dimensions of the fiber.

\section{The assumption of uniform liquid pressure for the 3D case}\label{AppendixE}
Next, we estimate the time scale for the fluid to equilibrate given nonuniform pressure distribution. To do so, we use Darcy's law
\begin{equation}
q = \frac{\kappa}{\mu}\nabla P
\label{Eq:Darcy}
\end{equation}
where $q$ is the flux (discharge per unit area, with units of length per time), $\kappa$ is the permeability of the two-dimensional pillar array, $\mu$ is the viscosity of the fluid, and $\nabla P$ is the pressure gradient. $\nabla P\sim \frac{\sigma}{D \ell}$, where $\sigma$ is the surface tension of the liquid, $D$ is the pillar spacing, and $\ell$ is the system size. $\kappa$ is approximated from a periodic square array of parallel cylinders, which is the initial configuration of our system. In the dilute solid volume fraction (porosity) regime, the asymptotic expression \cite{Sangani1982} for $\kappa$ is
\begin{equation}
\kappa = R^2\frac{-0.5\ln\phi-0.745+\phi-0.25\phi^2}{4\phi},
\label{Eq:kappa}
\end{equation}
where $R$ is the radius of the pillar, $\phi = \pi R^2/D^2$ is the porosity. In our typical experiment, $R=150 n m$, $L = 4.5\mu m$, $D=2\mu m$, $E=0.2GPa$, and $\sigma = 0.022N/m$. For a domain of size $\ell\sim100\mu m$, which contains thousands of pillars, the time scale for the fluid to relax so that the pressure inside the domain is uniform can be estimated as $\tau_f = \ell / q \sim 10^{-3}<\tau_b$.

In conclusion, for a reasonably large patch of pillars, it is plausible to assume that the pressure throughout the liquid is uniform during the evaporation because of the well-separated time scales -- the time scale for the fluid to relax ($\sim10^{-3}s$ for $50$ by $50$ pillars) $\ll$ that for the pillars to respond ($\sim10^{-2}s$) $\ll$ that for the evaporation ($\sim 10^0s$).

\section{Descriptions of movies}\label{AppendixF}
Movie 1: 2D Hierarchical\ Bundles\ Experiment.mov

A one-dimensional array of elastic lamellae made of polydimethylsiloxane (PDMS) with pre-polymer and cross-linker weight ratio 10:1 is immersed in the isopropyl alcohol (IPA) at room temperature $25^oC$. The lamellae have height $40\mu m$, thickness $10\mu m$, depth (into the screen) $40\mu m $, and uniform spacing $10\mu m$. The coalescence of lamellae initiates from the right and propagates to the left. Dimers appear first and further collapse into quadrimers. Bundles separate in the reverse order as the liquid evaporates. The surface adhesion force traps the final configuration in the dimer form. The movie is in real time and the duration is about $3s$.

\vspace{5mm}
\noindent Movie 2: 2D Irregular\ Bundles\ Experiment.mov

The material and geometric parameters of the lamella array are the same as those in Movie 1 except that the depth of the lamella is of order millimeters. The geometric imperfections of the system is larger than that in Movie 1. The coalescence initiates from multiple sites simultaneously and irregular bundles with size varying from 2 to 5, although the primary mode at the onset is still the dimer. The movie is in real time and the duration is about $34s$.

\vspace{5mm}
\noindent Movie 3: 2D Hierarchical\ Bundles\ Simulation.mov

A periodic array of 100 plates is simulated, and only 32 plates are displayed. The system starts with $\theta_n=\pi/2$, and $V_n/V_{flat}=0.92$ except that $V_{99}$ is $2\%$ less, where $V_{flat}$ is the control volume inside the system when all the pillars are vertical and the air-liquid interface is flat. The coalescence initiates from the $99^{th}$ cell and propagates in both directions. Dimers arise first and further collapse into quadrimers right after. The bundles separate in the reverse order as the liquid evaporates. The system recovers the configuration of uniform vertical lamellae in the end as no adhesion forces are included in the theoretical model. In order to see the onset of instability and hierarchical transitions clearly, the movie is not uniform in time, and the dimensionless time is included in the movie. The time scale is $\tau=22.5ms$, and the entire process completes in around $7s$ in real time.

\vspace{5mm}
\noindent Movie 4: 2D Irregular\ Bundles\ Simulation.mov

A periodic array of 100 plates is simulated, and only 31 plates are displayed. The initial condition is the uniform random perturbations with the maximum amplitude of $5\%$ applied on both $V_n/V_{flat}= 0.92$ and $\theta_n=\pi/2$, where $V_{flat}$ is the control volume inside the system when all the pillars are vertical and the air-liquid interface is flat. The coalescence initiates from multiple sites simultaneously. The dominant onset mode is still the dimer. During the process of evaporation, irregular bundles of size varying from 2 to 5 form and eventually separate. In order to see the onset of instability and early stages of coalescence clearly, the movie is not uniform in time, and the dimensionless time is included in the movie. The time scale is $\tau=22.5ms$, and the entire process completes in around $7s$ in real time.

\vspace{5mm}
\noindent Movie 5: 3D$\_$14x14$\_$V090.mov

It shows the simulated evolution of an array of $14$ by $14$ pillars collapsing into fourfold bundles for a prescribed liquid volume of $V/V_{flat}=0.90$, where $V_{flat}$ is the control volume inside the system when all the pillars are vertical and the air-liquid interface is flat, using the parameters from experiments in Figure 1(c)-(e). The final stable configuration at the end of this movie corresponds to Figure 7(a).

\vspace{5mm}
\noindent Movie 6: 3D$\_$10x16$\_$V088.mov

It shows the simulated evolution of an array of $10$ by $16$ pillars collapsing into fourfold bundles for a prescribed liquid volume of $V/V_{flat}=0.88$, where $V_{flat}$ is the control volume inside the system when all the pillars are vertical and the air-liquid interface is flat, using the parameters from experiments in Figure 1(c)-(e). The final stable configuration at the end of this movie corresponds to Figure 7(b).

\bibliographystyle{prs}

\end{document}